\documentclass[final,journal,twocolumn,12pt]{IEEEtran}
\IEEEoverridecommandlockouts
\usepackage{epsfig,amsmath,amssymb,amsfonts,bm}
\usepackage{cite}
\usepackage{color}
\usepackage{soul}
\usepackage{subfigure}
\usepackage{epstopdf}
\usepackage{overpic}
\graphicspath{{figs/}}
\usepackage{booktabs}
\usepackage{rotating}
\usepackage{stfloats}
\usepackage[colorlinks=false,
linkcolor=red,
anchorcolor=blue,
citecolor=green
]{hyperref}
\allowdisplaybreaks[4]

\usepackage{rotating}
\usepackage{array}
\usepackage{multirow}
\usepackage{longtable}
\usepackage{booktabs}
\usepackage{newtxtext, newtxmath}
\begin{document}

\title{Principal Component Maximization: A Novel Method for SAR Image Formation from Raw Data without System Parameters}

\author{
Huizhang~Yang,
Zhong~Liu, and
Jian~Yang
\thanks{
This work was supported in part by Natural Science Foundation of China under grant No. 62301259 and 62171224, the Natural Science Foundation of Jiangsu Province, China, under grant No.BK20221486, and the Fundamental Research Funds for the Central Universities under grant No.30924010914.

Huizhang Yang and Zhong Liu are with the School of Electronic and Optical Engineering, Nanjing University of Science and Technology, Nanjing 210094, China (e-mail: hzyang@njust.edu.cn)

Jian Yang is with the Department of Electronic Engineering, Tsinghua University, Beijing
100084, China.
}
}
\maketitle
\begin{abstract}

Synthetic aperture radar (SAR) imaging traditionally requires precise knowledge of system parameters to implement focusing algorithms that transform raw data into high-resolution images. These algorithms require knowledge of SAR system parameters, such as wavelength, center slant range, fast time sampling rate, pulse repetition interval (PRI), waveform parameters (e.g., frequency modulation rate), and platform speed.
This paper presents a novel framework for recovering SAR images from raw data without the requirement of any SAR system parameters. Firstly, we introduce an approximate matched filtering model that leverages the inherent shift-invariance properties of SAR echoes, enabling image formation through an adaptive reference echo estimation. To estimate this unknown reference echo, we develop a principal component maximization (PCM) technique that exploits the low-dimensional structure of the SAR signal.  The PCM method employs a three-stage procedure: 1) data block segmentation, 2) energy normalization, and 3) principal component energy maximization across blocks, effectively handling non-stationary clutter environments. Secondly, we present a range-varying azimuth reference signal estimation method that compensates for the quadratic phase errors. For cases where PRI is unknown, we propose a two-step PRI estimation scheme that enables robust reconstruction of 2-D images from 1-D data streams.  Experimental results on various SAR datasets demonstrate that our method can effectively recover SAR images from raw data without any prior system parameters.

\end{abstract}
\begin{IEEEkeywords}
Synthetic aperture radar (SAR), SAR imaging, auto focus, principal component maximization
\end{IEEEkeywords}
\IEEEpeerreviewmaketitle

\section{Introduction}\label{s1}

Synthetic Aperture Radar (SAR) operates by sequentially emitting pulses of microwave signals towards a target area and recording the echoes scattered from that area. The recorded raw data over $N$ pulse observation intervals are organized into 2-D datasets, which are then used to form the radar image of the illuminated scene. The image formation process utilizes these 2-D datasets to coherently accumulate raw data for each 2-D resolution cell, a procedure known as SAR focusing\cite{cumming2005digital}. This focusing process relies on various SAR system parameters. Numerous algorithms have been developed for implementing SAR focusing, primarily falling into two categories: frequency-domain algorithms and time-domain algorithms.

\subsection{SAR Focusing Algorithms}

Frequency-domain algorithms, including the Range-Doppler Algorithm (RDA), Chirp Scaling Algorithm (CSA), and Range Migration Algorithm (RMA), offer computational efficiency and are widely utilized in practical systems. The distinction among these algorithms lies in their approach to correcting range cell migration (RCM), which reflects the range-azimuth coupling of SAR echoes. In RDA, targets with the same slant range but different azimuth cells share the same RCM trajectory in the range-Doppler domain. Correcting this trajectory in the domain is equivalent to correcting a set of target trajectories at the slant range, achievable through interpolation \cite{cumming2005digital}. In CSA, RCM correction involves multiplying by an exponential phase term in the 2-D frequency domain to correct bulk RCM, followed by correction of residual RCM in the fast time and Doppler domain by exploiting the quadratic phase of chirp signals \cite{raneyPrecisionSARProcessing1994}. For RMA, RCM correction is performed by multiplying with a reference function in the 2-D frequency domain \cite{cafforioSARDataFocusing1991}. Both RDA and RMA require interpolation, necessitating a balance between interpolation precision and computational load. Another important concept reflecting range-azimuth coupling is the additional second-order phase term in range time. Correcting this term, known as second-order range correction, depends on the specific algorithm used.

The standard versions of the aforementioned algorithms are traditionally applied to stripmap mode data. For other imaging modes, such as spotlight and TOPS, efficient image focusing requires adding pre- and/or post-processing steps tailored to the specific needs of each SAR mode to address issues like spectrum folding and rotation \cite{dezanTOPSARTerrainObservation2006, pratsProcessingSlidingSpotlight2010a, lanariSpotlightSARData2001b, pratsVeryHigh2014, zhuExtendedTwoStep2021}. Some algorithms have been developed specifically for certain imaging modes. PFA is an imaging algorithm designed for the spotlight mode. This method demodulates the 2-D signal using a reference linear frequency modulation (LFM) signal, compensates for the residual video phase, performs 2-D interpolation, and then applies a 2-D Fourier transform to obtain images. However, due to the assumption of a plane wavefront, it often requires compensation for wavefront curvature errors in very high-resolution images \cite{maoTwodimensionalAutofocusSpotlight2016}.

Sun et al. proposed a unified focusing algorithm based on the Fractional Fourier Transform (FrFT), capable of processing data from stripmap, spotlight, sliding spotlight, and TOPS modes \cite{sunUnifiedFocusingAlgorithm2013}. The core idea is to use FrFT to unfold the spectrum in either the time or frequency domains and reduce the data length to improve processing efficiency.

Another class of SAR focusing algorithms includes time-domain algorithms, such as the Back Projection (BP) algorithm and its variants. BP \cite{munson983} uses precise signal expressions and imposes no restrictions on platform motion trajectories, facilitating motion error compensation and enabling high-resolution image generation. BP works by projecting range-compressed echo data into the image domain via coherent summation along the trajectory of each resolution cell. Fast Back Projection (FBP) \cite{yegulalp1999a} and Fast Factorized Back Projection (FFBP) \cite{ulander2003a} accelerate BP by first generating sub-aperture low-resolution images using the low angular bandwidth in sub-aperture polar coordinates, then coherently merging them to form high-resolution images. However, the requirement for interpolation during the coherent merging process still poses significant computational burdens. To mitigate this limitation, the Cartesian Factorized Back Projection Algorithm (CFBP) \cite{liangFastTimeDomainSAR2019} was developed to generate and merge sub-aperture images using Cartesian coordinates. The sub-aperture images have folded spectra, but the application of an azimuth reference function can deramp or compress the azimuth spectra, allowing for merging without interpolation.

\subsection{SAR Autofocus Algorithms}

For airborne SAR systems, flight trajectories often deviate from the ideal due to factors such as atmospheric disturbances and imprecise flight control. These deviations introduce errors in the observed signals, leading to defocusing in SAR images. Consequently, additional processing known as motion compensation (MoCo) is necessary. Two classes of MoCo methods exist. The first involves reconstructing motion errors using data from inertial navigation systems (INS) and/or the global positioning system (GPS), and then compensating for these errors in the signal's envelope and phase in specific domains \cite{kennedy1988, moreira1990, buckreuss1994}. When INS/GPS systems are either unavailable or lack the precision required for high-resolution imaging, data-driven estimation and compensation of motion-induced errors (or residual errors after MoCo using INS/GPS data) become essential; this process is referred to as autofocus \cite{chen2022a}. Numerous autofocus algorithms have been developed.

The MapDrift (MD) algorithm \cite{samczynski2010} estimates polynomial phase errors by analyzing the mutual drift of two or more sub-aperture images. Phase Gradient Autofocus (PGA) \cite{wahl1994a} estimates motion errors based on the phase gradient of dominant scatterers. Image quality optimization methods, which use criteria such as maximum contrast/sharpness or minimum entropy, estimate motion errors as nonparametric vectors or polynomials through iterative searches \cite{lixi1999a, berizzi2002, gao2014, zhang2015a, zeng2016}. These methods are designed to address space-invariant errors. For space-variant errors, data can be split into multiple range blocks, with each block's motion error estimated and fitted using certain criteria, such as weighted least squares \cite{chen2022a}. The estimated motion errors can then be compensated for through additional steps within focusing algorithms.

In the SAR focusing algorithms mentioned above, knowledge of SAR system parameters is essential for generating images. In scenarios where these parameters are unknown, such as when they are unavailable or when the SAR illuminating source is non-cooperative, then these algorithms cannot be used to generate SAR images.


\subsection{The Solution and Contributions of This Paper}

Motivated by the aforementioned issues, this paper investigates the problem of recovering SAR images from raw SAR data without the requirement of any SAR system parameters. 
Our solution and contributions to the above problem are summarized as follows.

\begin{itemize}
    \item Development of a novel model for SAR image representation that does not require any prior knowledge of SAR system parameters. This model is based on the approximate shift-invariance property of point echoes and uses an unknown reference echo for matched filtering.
    \item Development of a Principal Component Maximization (PCM) method to estimate the reference echo from raw 2-D data. The PCM method segments the echo data into blocks, normalizes each block's energy, and estimates the reference echo by maximizing the principal component's energy across all blocks, effectively suppressing the impact of non-stationary clutter.
    \item Development of a two-step approach for recovering the 2-D raw data matrix from the 1-D raw data vector when the pulse repetition interval (PRI) is unknown. This approach utilizes the approximate periodicity of amplitude signals and PCM for coarse and fine PRI estimation, enabling accurate transformation of 1-D data into a 2-D matrix.
    \item Development of a range-dependent azimuth reference signal estimation method to address quadratic phase errors introduced by the approximated shift-invariance model. This method compensates for these errors, significantly improving image quality in the azimuth direction.
\end{itemize}

The effectiveness and robustness of the proposed method are validated through extensive experiments using ERS, RADARSAT, and airborne SAR datasets.

The remainder of this paper is organized as follows. Section \ref{sec:sg-model} builds signal models for a reference point's echo and clutter and analyzes the low-dimensional structure of point echoes. Section \ref{sec:img-rp} develops an image representation model for SAR using the reference point's echo via an approximated shift-invariance model of point echoes. Section \ref{sec:est-Yp} presents our PCM method for estimating the reference point's echo. Section \ref{sec:est-T} describes a two-step method for recovering 2-D SAR data from 1-D data without knowledge of the PRI. Section \ref{sec:err-correct} analyzes phase error and provide a correction method. Section \ref{sec:exp} provides experimental validation. Finally, Section \ref{sec:conc} concludes the paper.

\section{Signal Model}\label{sec:sg-model}

The signal model for SAR data is a fundamental component of our analysis. It provides a mathematical representation of the echoes received by the SAR system and is essential for understanding the underlying principles of SAR imaging. The models presented in this section form the basis for the image recovery algorithm developed in subsequent sections.

\subsection{Point Target Echo Model}

We begin our analysis by considering a reference point $(x_0, R_0)$ in the SAR azimuth-slant range plane. The coordinates of this point in the slow-fast time domain are expressed as $p = [\eta_0, \tau_0] = [x_0/v, 2R_0/c]$, where $v$ is the platform velocity, and $c$ is the speed of light. The two-dimensional echo from this point can be represented by the equation
\begin{equation}
\begin{split}
Y_p(\eta,\tau) = &a_p w_{\rm a}(\eta-\eta_0) s\left(\tau - \frac{2R(\eta,p)}{c}\right) \\
&\exp\left(-j4\pi \frac{R(\eta,p)}{\lambda}\right)
\end{split}
\end{equation}
Here, $w_{\rm a}(\eta)$ denotes the azimuth antenna pattern weighting function, $s(\tau)$ is the baseband of the transmitted pulse signal, and $R(\eta,p)$ is the range history. $\lambda$ represents the radar waveform length. It is important to note that while these parameters are used to model the SAR echo, the image recovery algorithm developed in this paper does not require any knowledge of these parameters. The range history $R(\eta,p)$ is dependent on the specific transmit and receive geometry. For a monostatic configuration, it can be approximated as
\begin{equation}
R(\eta,p) = \sqrt{(v\eta-x_0)^2+R_0^2} \approx R_0 + \frac{(v\eta-x_0)^2}{2R_0}
\end{equation}

\subsection{Low-Dimensional Model for Point Target Echo}

\begin{figure}
\centering
\includegraphics[width=7cm]{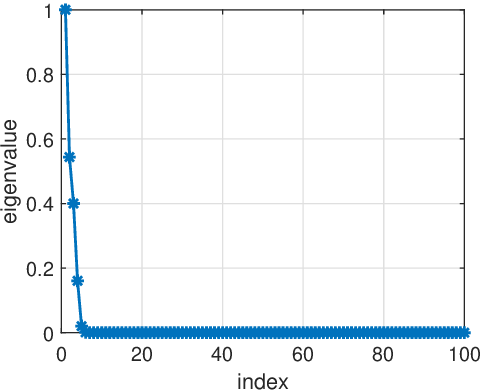}
\caption{Top 100 eigenvalues of a simulated $2048 \times 2048$ echo of a point target.}\label{fig:eigen_value}
\end{figure}

The point echo can be approximated as the sum of a few azimuth-range decoupled signals, which implies a low-dimensional structure. This property is pivotal in the development of our image recovery algorithm, which will be detailed in the subsequent sections. To demonstrate this property, we perform a two-dimensional order-$n$ Taylor expansion on the point echo
\begin{equation}\label{eq:Yp}
\begin{split}
&Y_p(\eta,\tau) \\
&\approx Y_p(\eta_0,\tau_0)+\left.{\sum_{p=1}^{n}\left(\eta\frac{\partial }{\partial\eta}+\tau\frac{\partial}{\partial\tau}\right)^p Y_p(\eta,\tau)}\right|_{(\eta_0,\tau_0)}
\end{split}
\end{equation}
The second term of this expansion can be further simplified to
\begin{equation}\label{}
\sum_{p=1}^{n}\sum_{q=0}^{p}\binom{p}{q}\left(\eta-\eta_0\right)^q \left(\tau-\tau_0\right)^{p-q}\frac{\partial^p Y_p(\eta,\tau)}{\partial \eta^q\partial \tau^{p-q}}\bigg |_{(\eta_0,\tau_0)}
\end{equation}
We observe that this equation is a summation of a sequence of range-azimuth decoupled signals, which can be reexpressed as
\begin{equation}\label{}
\rho_k u_k(\eta) v_k(\tau), \quad k=1,\cdots,K
\end{equation}
where $K=\sum_{p=1}^{n}\sum_{q=0}^{p}\binom{p}{q}$. Since the rank of $z_k a_k(\eta) b_k(\tau)$ in the above equation is $1$, as it is range-azimuth decoupled, the order-$n$ Taylor expansion in equation in (\ref{eq:Yp}) has a rank not larger than $K$+1. For example, for order-$3$ Taylor expansion, the rank is not larger than $15$. We refer to this property as the low-dimensional property of point echo. This property allows us to express the point echo using its SVD
\begin{equation}
Y_p(\eta,\tau) \approx \sum_k^K \rho_k u_k(\eta) v_k(\tau)
\end{equation}
where $K$ is a small number. Here $\rho_k$ are the singular values and can be obtained by SVD on the echo data matrix of eigendecomposition on the echo covariance matrix. The low-dimensional property can be visually analysed from the eigenvalues $\rho_k^2$ calculated simulated point echo data matrix. Fig. \ref{fig:eigen_value} provides such an example of calculated top 100 eigenvalues from a simulated broadside stripmap SAR data matrix. From this figure, we can find that there only exist a few large eigenvalues, and most eigenvalues are negligible. In particular, the first eigenvalue is significantly than the others.

\subsection{Clutter Model}

In this study, we consider the presence of a strong point scatterer, which serves as our reference point, while the remainder of the scatterers are treated as background clutter. The echo from these clutter cells is modeled as follows.

Consider clutter cells located at $(x_{c_k}, y_{c_k})$, for $k=1,\cdots,K_c$. Their coordinates in the 2-D time domain are given by $c_k = [\eta_{c_k}, \tau_{c_k}] = [x_{c_k}/v, 2R_{c_k}/c]$. The clutter scattering from these cells can be expressed as
\begin{equation}
\begin{split}
Y_c(\eta,\tau) =\sum_k &a_{c_k} w_{\rm a}\left(\eta-\frac{x_{c_k}}{v}\right) s\left(\tau - \frac{2R(\eta,c_k)}{c}\right) \\
&\cdot\exp\left(-j4\pi \frac{R(\eta,c_k)}{\lambda}\right)
\end{split}
\end{equation}
where
\begin{equation}
R(\eta,c_k) = \sqrt{(v\eta-x_{c_k})^2+R_{c_k}^2}
\end{equation}
Clutter is essentially the superposition of echoes from various scattering elements, with each element's echo amplitude being random. The intensity of clutter is influenced by factors such as the type of ground objects, the incidence angle, and the beam pattern. On a global scale, clutter is considered non-stationary. However, on a local scale, the intensity of clutter is approximately stationary and follows an independent and identically distributed pattern.

\subsection{One-Dimensional Observation Model}

The observed signal is a composite of the point echo, clutter, and noise, which can be expressed as
\begin{equation}
Y(\eta,\tau) = Y_p(\eta,\tau) + Y_c(\eta,\tau) + Y_n(\eta,\tau)
\end{equation}
where $Y_n(\eta,\tau)$ denotes the noise term.

It is important to note that SAR signals are actually observed in a 1-D time domain. The 2-D representation provided here is a rearrangement of the 1-D data with a time interval $T$ (the Pulse Repetition Interval, PRI). The relationship between the 1-D time $t$ and the 2-D time $(\eta,\tau)$ is given by
$t = \eta + \tau$, where the slow time $\eta$ takes discrete values $\eta = nT$, for $n=1,\cdots,N$. The mapping from the 1-D signal $y(t)$ to the 2-D domain can be expressed as
\begin{equation}
Y(\eta,\tau) = \mathcal{M}_{\rm 1d\rightarrow 2d}\left\{y(t);T\right\}
\end{equation}

The inclusion of this 1-D to 2-D mapping is to emphasize that the PRI parameter $T$ is necessary for forming 2-D data in conventional SAR image focusing. However, in this paper, we do not assume any prior knowledge of SAR operating parameters, including $T$. Instead, we will estimate this parameter from the 1-D data, and the estimation method will be presented in Section \ref{sec:est-T}.

\section{Image Representation Model via Approximated Shift-Invariance}\label{sec:img-rp}

In this section, we propose an approximated matched filtering model to describe the SAR image using the 2-D observation signal $Y(\eta,\tau)$. This model will form the basis for developing our SAR image recovery algorithm, which will be detailed in Section \ref{sec:est-Yp}.

\subsection{Approximated Shift Invariance of Point Target Echo}

Let us consider a generic point located at a distance $\Delta p$ from the reference point. The coordinates of this generic point are given by
$p + \Delta p = [\eta_0 + \Delta \eta, \tau_0 + \Delta \tau] = [(x_0 + \Delta x)/v, 2(R_0 + \Delta R)/c]$.

The transformation of this generic point $p + \Delta p$ into its corresponding 2-D SAR echo can be represented by a system function $H_{p+\Delta p}$. Although this function varies with $p+\Delta p$, we will assume that it is invariant with respect to $p+\Delta p$. The following development will show the modeling process.

\begin{figure}
\centering
\includegraphics[width=6cm]{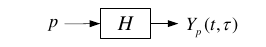}
\caption{Approximated system model mapping a point target $p$ to its echo $Y_p$ using a shift-invariant system $H$.}\label{fig:H}
\end{figure}

Using a second-order approximation of the range history $R(\eta,p+\Delta p)$, we obtain
\begin{equation}
\begin{split}
R(\eta,p+\Delta p) \approx & R_0 + \Delta R + \frac{(v\eta-x_0-\Delta x)^2}{2R_0}\\
\approx & \Delta R + \sqrt{(v\eta-x_0-\Delta x)^2 + R_0^2} \\
= & \Delta R + R(\eta-\Delta x / v,p)
\end{split}
\end{equation}
This equation suggests that the range history of a generic point can be approximated as the range history of the reference point plus the range difference $\Delta R$. With this approximation, the echo of $p+\Delta p$ can be expressed as a shifted version of the reference echo. The following derivation illustrates this relationship
\begin{equation}
\begin{split}
Y_{p+\Delta p}(\eta,\tau) = &a_p w_{\rm a}(\eta-(x_0+\Delta x)/v) \\
&\cdot s(\tau - 2R(\eta,p+\Delta p)/c) \\
&\cdot \exp(-j4\pi R(\eta,p+\Delta p)/\lambda)\\
\approx &a_p w_{\rm a}(\eta-(x_0+\Delta x)/v) \\
&\cdot s(\tau - 2\Delta R/c - 2R(\eta-\Delta x / v,p)/c) \\
&\cdot \exp(-j4\pi R(\eta-\Delta x/v,p)/\lambda) \\
&\cdot \exp(-j4\pi \Delta R/\lambda)\\
\approx &Y_p(\eta-\Delta x / v, \tau-2\Delta R / c) \\
&\cdot \exp(-j4\pi \Delta R/\lambda)
\end{split}
\end{equation}
By ignoring the exponential phase term containing $\exp(-j4\pi \Delta R/\lambda)$, we can treat the point echo response as a shift-invariant one with respect to $p$, i.e.,
\begin{equation}
Y_{p+\Delta p}(\eta,\tau) \approx Y_p(\eta-\Delta \eta, \tau-\Delta \tau)
\end{equation}
Accordingly, the mapping of point scatterer to its echo can be approximately modelled as a shift-invariant system $H$ as illustrated in Fig. \ref{fig:H}. This property will be used to model SAR images for developing our method.

\subsection{Image Representation with Reference Echo Based on the Approximated Shift Invariance}

SAR image focusing can be conceptualized as a process that focuses the echo from each resolution element to form a complex image. The amplitude of this image represents the focused echo intensity, while the phase contains range information relative to a reference point. The focusing process can be described as follows.

For the reference point $p$, the focused pixel value is given by the inner product
\begin{equation}
\begin{split}
I(p) = &\iint Y_p(\eta,\tau)\,Y_p^*(\eta,\tau) \,d\eta d\tau\\
\approx &\iint Y(\eta,\tau)\,Y_p^*(\eta,\tau) \,d\eta d\tau
\end{split}
\end{equation}
where $(\cdot)^*$ denotes the conjugate operation.

For an arbitrary point $p+\Delta p$, the focused amplitude is represented by $\iint Y_{p+\Delta p}(\eta,\tau)\,Y_{p+\Delta p}^*(\eta,\tau)\,d\eta d\tau$, and the phase term that characterizes the range difference with respect to the reference point is given by $\exp(-j4\pi \Delta R/\lambda)$. Therefore, the focused intensity at point $p+\Delta p$ can be expressed as
\begin{equation}
\begin{split}
&I(p+\Delta p) \\
&= \exp(-j4\pi \Delta R/\lambda) \iint Y_{p+\Delta p}(\eta,\tau)\, Y_{p+\Delta p}^*(\eta,\tau)\,d\eta d\tau
 \\
&\approx \iint Y_{p+\Delta p}(\eta,\tau) \,Y_p^*(\eta-\Delta x / v, \tau-2\Delta R / c)\,d\eta d\tau \\
&\approx\iint Y(\eta,\tau)\, Y_p^*(\eta-\Delta x / v, \tau-2\Delta R / c) \,d\eta d\tau
\end{split}
\end{equation}

By discretizing the observed scene using grids
\begin{equation}
\Delta p[n,k] = \left[n\,\rho_x/v, k\,2\rho_R/c\right]
\end{equation}
where $\rho_x$ is the range resolution and $\rho_R$ is the azimuth resolution, the SAR image can be represented as the 2-D focused responses of all the grid points:
\begin{equation}
\begin{split}
I[n,k] =& I(p + \Delta p[n,k]) \\
\approx & \iint Y(\eta,\tau) \,Y_p^*(\eta-n\Delta_{\eta}, \tau-k\Delta_{\tau})\, d\eta d\tau
\end{split}
\end{equation}\label{eq:mf}
Here, $\Delta_{\eta} = \rho_x/v$ is the slow-time sampling interval, and $\Delta_{\tau} = 2\rho_R/c$ is the fast-time sampling interval. Since the sampled data is discrete, the above equation can be rewritten in a discrete form
\begin{equation}
\begin{split}
I[n,k] \approx & \sum\limits_{i,j} Y(i\Delta_{\eta},j\Delta_{\tau})\,Y_p^*(i\Delta_{\eta}-n\Delta_{\eta}, j\Delta_{\tau}-k\Delta_{\tau}) \\
= & \sum\limits_{i,j} Y[i,j]\,Y_p^*[i-n,j-k] \\
= & Y[n,k] \circ Y_p^*[-n,-k]
\end{split}
\end{equation}

Based on the above image formation model, we can recover a SAR image by correlating the 2-D echo with a reference echo. This process can be efficiently implemented using fast convolution, which is a computationally efficient method for convolving two signals.

Fast convolution leverages the Fast Fourier Transform (FFT) to perform the convolution in the frequency domain, which is particularly useful for large datasets such as SAR images. By transforming both the observed signal and the reference echo into the frequency domain, multiplying them, and then inverse transforming the result back into the spatial domain, we can obtain the focused SAR image. This approach significantly reduces the computational complexity compared to direct convolution in the time domain.

In the next section, we will detail the algorithm for estimating the reference echo $Y_p(\eta,\tau)$ and demonstrate how it can be used to recover the SAR image from the 1-D raw radar data.

\subsection{Properties of the Image Representation Model}

Since the reference echo is used for matched filtering, the reference point is perfectly focused in this model. This suggests that range compression, RCMC, and azimuth compression are simultaneously performed for the reference point. Conventionally, RCMC and azimuth compression in SAR focusing algorithms are range-varying, but they are fixed for all range in our model under the shift-invariance approximation. Accordingly, there exist residual RCMC and azimuth defocusing in our model.

It is important to note that the reference echo $Y_p$ is not known a priori and must be estimated from the available data.  The estimation of the reference echo can significantly affect the quality of the recovered image, and thus, it is the key step in our image recovery algorithm.

However, this estimation is a challenging task due to two problems: 1) we only have the raw echo data and we cannot directly select reference point in the echo data domain. Thus, we need to directly estimate the echo of the reference point in the echo data domain, which often have strong and non-stationary clutter. 2) without any system parameters including the PRI $T$, the raw data is 1-D. We need to develop a data-driven method for transforming the 1-D raw data to its 2-D version.
The solution to the above problems are developed in sections \ref{sec:est-Yp} and \ref{sec:est-T}, respectively.


\section{Image Recovery via Principal Component Maximization (PCM)}\label{sec:est-Yp}

Building upon the previous image formation model, this section introduces a novel approach to estimate the reference point echo and subsequently recover the SAR image using the previous representation model in (\ref{eq:mf}). Our method involves segmenting the echo data into multiple blocks, estimating the reference echo by maximizing the principal component across all blocks, and finally recovering the image through the convolution in (\ref{eq:mf}).

\subsection{Block Segmentation and Normalization of 2-D Raw Data}

\begin{figure}
\centering
\includegraphics[width=8cm,height=6cm]{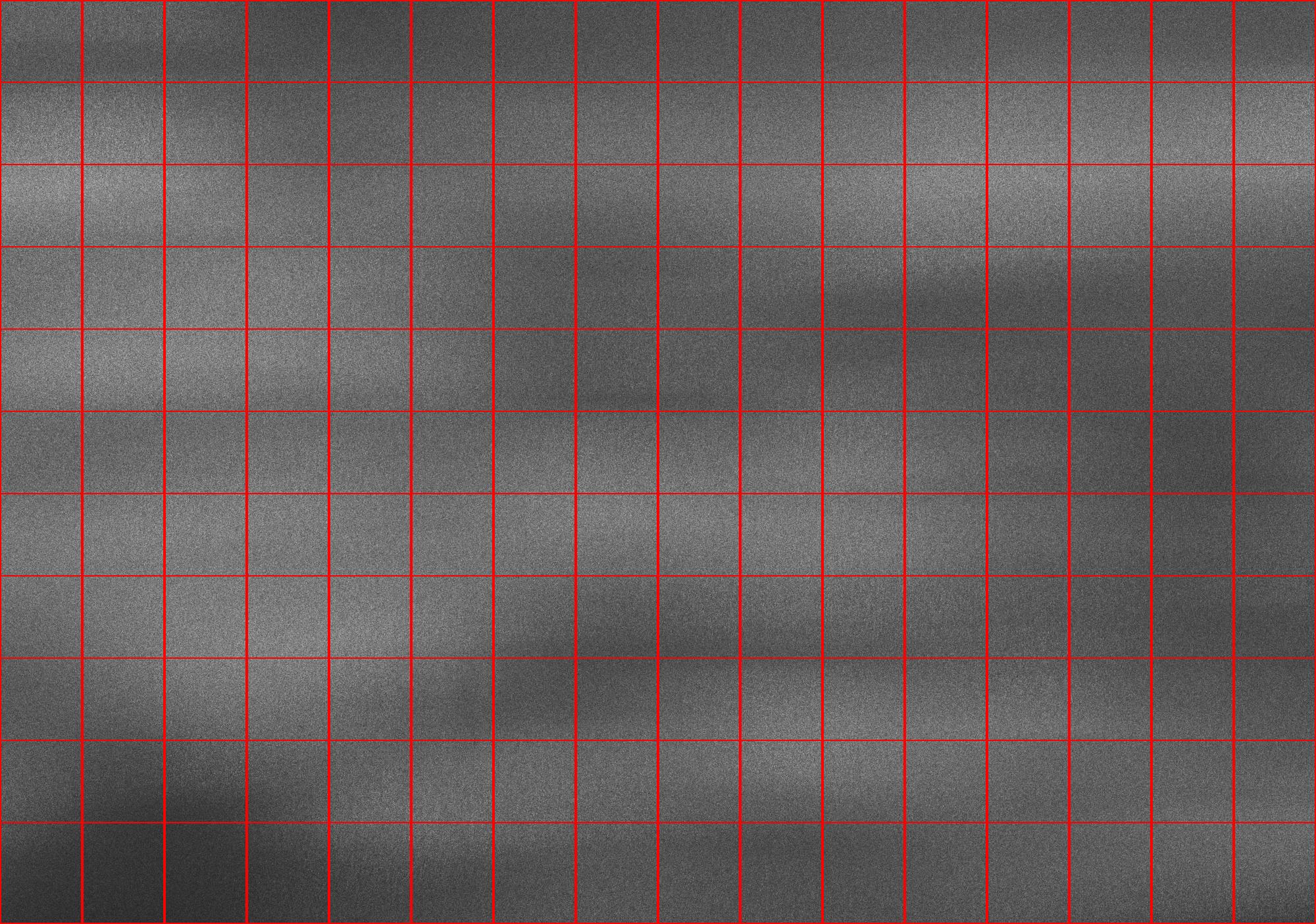}
\caption{Illustration of block segmentation. ERS raw data \url{E2_16385_STD_L0_F370} is used in this example.}\label{fig:block_seg}
\end{figure}

To effectively estimate the reference echo in the presence of clutter, we must account for the time-domain characteristics of clutter. Clutter is the summed echoes from various background scattering elements, and its intensity is influenced by factors such as ground object type, incidence angle, and beam pattern. Clutter is generally considered non-stationary in a global sense. To address the non-stationarity of clutter, we segment the observed data into blocks and normalize the Frobenius norm of each block. Fig. \ref{fig:block_seg} provides a visual demonstration of the block segmentation process.
The normalization process for $k$-th raw data block $\mathbf{Y}_k$ can be expressed as
\begin{equation}\label{eq:nm}
  \mathbf{Y}_{k,\rm nm} = \frac{\mathbf{Y}_k}{\|\mathbf{Y}_k\|_F}
\end{equation}
The normalization is to equalize the energy of each data block so that the influence of non-stationary clutter among difference blocks can be suppressed in the estimation of reference echo via PCM.

\begin{figure}
\centering
\includegraphics[width=8.8cm]{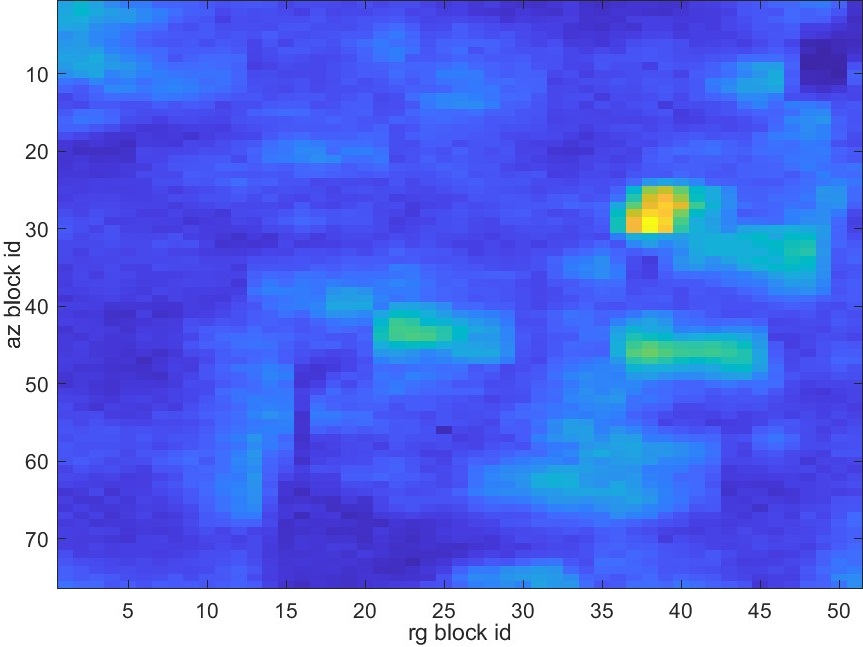}
\caption{Illustration of PCM across $76\times 51$ raw data blocks. The brightest point indicates the block that maximizes objective function in the PCM formulation (\ref{eq:p1}). ERS raw data \url{E2_16385_STD_L0_F370} is used in this example.}\label{fig:block_seg2}
\end{figure}

\subsection{Reference Echo Recovery via Principal Component Maximization}

The reference echo is estimated by PCM which maximizes the principal component across all normalized raw data blocks. This optimization problem can be formulated as follows
\begin{equation}\label{eq:p1}\tag{P1}
\begin{array}{lll}
\hat{\mathbf{Y}}_{\rm ref} =
& \mathop{\rm arg max}\limits_{\mathbf{Y}_{k,\rm pc}} &\inf\{g_k | k=1,\cdots,K\}\\
&\hbox{subject to} &g_k = \|\mathbf{Y}_{k,\rm pc}\|_F^2\\
&                             &\mathbf{Y}_{k,\rm pc}={\cal P}(\mathbf{Y}_{k,\rm nm},D)
\end{array}
\end{equation}
Here, $\mathbf{Y}_{k,\rm pc}={\cal P}(\mathbf{Y}_{k,\rm nm},D)$ represents the $D$-dimensional principal component of the $k$-th normalized raw data block $\mathbf{Y}_{k,\rm nm}$, and can be expressed as the solution of the following rank-$D$ approximation problem
\begin{equation}\label{eq:p2}\tag{P2}
\begin{array}{lll}
\mathbf{Y}_{k,\rm pc}=
&\mathop{\rm arg min}\limits_{\mathbf{V}} &\|\tilde{\mathbf{Y}}_{k,\rm nm} - \mathbf{V}\|_F^2\\
&\hbox{subject to} &{\rm rank}(\mathbf{V}) = D
\end{array}
\end{equation}
Observing  the problem (\ref{eq:p1}), we can find that $g_k = \|\mathbf{Y}_{k,\rm pc}\|_F^2$ is equivalent to the sum of the largest-$D$ eigenvalues $\sum_{d=1}^{D}\rho_d$ of the data covariance matrix $\mathbb{E} \{\mathbf{Y}_k \mathbf{Y}^{\rm H}_k \}$. Therefore, the physical concept of optimization problem (\ref{eq:p1}) is to maximize the principal component among all normalized blocks.

Given that the mean component of the data is removed, this optimization problem is effectively solved by
estimating the principal components $\mathbf{Y}_{k,\rm pc}$ for all normalized blocks and selecting $\mathbf{Y}_{k',\rm pc}$ having largest Frobenius norm (i.e., $\|\mathbf{Y}_{k',\rm pc}\|_F\ge \|\mathbf{Y}_{k,\rm pc}\|_F, \, \forall k$) as the echo, i.e., estimate $\hat{\mathbf{Y}}_{\rm ref} = \mathbf{Y}_{k',\rm pc}$.
Fig. \ref{fig:block_seg2} provides an example of the PCM across the $76\times 51$ raw data blocks.

\subsection{Selection of The Principal Component Dimension via Factorized Signal-to-Clutter-plus-Noise Ratio (SCNR) Analysis}

The SVD decomposition of the reference echo is approximately $\mathbf{Y}_{\rm ref} \approx \mathbf{U}{\rm diag}\{\sqrt{\rho_1},\cdots,\sqrt{\rho_D}\}\mathbf{V}$. The eigendecomposition of the reference echo's covariance matrix is
\begin{equation}\label{}
\mathbf{R}_{\rm ref} = \mathbf{Y}_{\rm ref} \mathbf{Y}_{\rm ref}^{\rm H} = \mathbf{U}{\rm diag}\{\rho_1,\cdots,\rho_D\}\mathbf{U}^{\rm H}
\end{equation}
Due to the presence of clutter and noise in the observed signal, we must consider their impacts on the principal component estimation. The eigendecomposition of each raw data block covariance matrix (consisting of point echo, clutter, and noise) is
\begin{equation}\label{}
\mathbf{R}_k  = \mathbf{Y}_k \mathbf{Y}^{\rm H}_k = \mathbf{U}\,{\rm diag}\{\rho_1,\cdots,\rho_D\}\mathbf{U}^{\rm H} + (\sigma_c + \sigma_n) \mathbf{I}
\end{equation}
The Signal-to-Clutter-plus-Noise Ratio (SCNR) of the $d$-th component is $\alpha_d = \rho_d/(\sigma_c + \sigma_n)$.
The energy of the reference echo $\mathbf{Y}_{\rm ref}$ is primarily concentrated in the first component, so the first eigenvalue is significantly larger than the others, i.e., $\rho_1 \gg \rho_2 > \cdots > \rho_D$. Consequently, the SCNR of the first component is significantly higher than those of the rest components, i.e., $\alpha_1 \gg \alpha_2 > \cdots > \alpha_D$. Therefore, we propose to use only the first component for estimating the reference echo.

\section{2-D Data Matrix Recovery from 1-D Raw Data Vector  with Unknown PRI}\label{sec:est-T}

The sample data stream of SAR echo is initially 1-D raw data. Its formation as 2-D raw data can be achieved only when the sample's PRI, $P= T f_{\rm s}$, is known. When this parameter is unknown, it must be estimated from the data, to reconstruct the 2-D raw data for the PCM and image formation processing mentioned in the previous section. To this end, in this section we propose a two-step method for estimating the sample PRI and recovering the 2-D raw data matrix from the 1-D raw data.

\subsection{Parametric Model of Transforming 1-D Raw Data to 2-D Raw Data}

\begin{figure*}
\centering
\subfigure[]{\includegraphics[width=16cm]{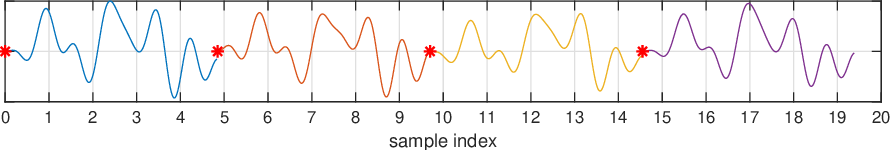}}\\
\subfigure[]{\includegraphics[width=7.7cm]{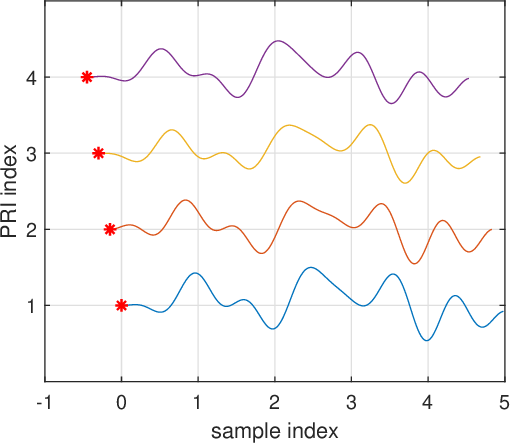}}
{\includegraphics[width=0.8cm]{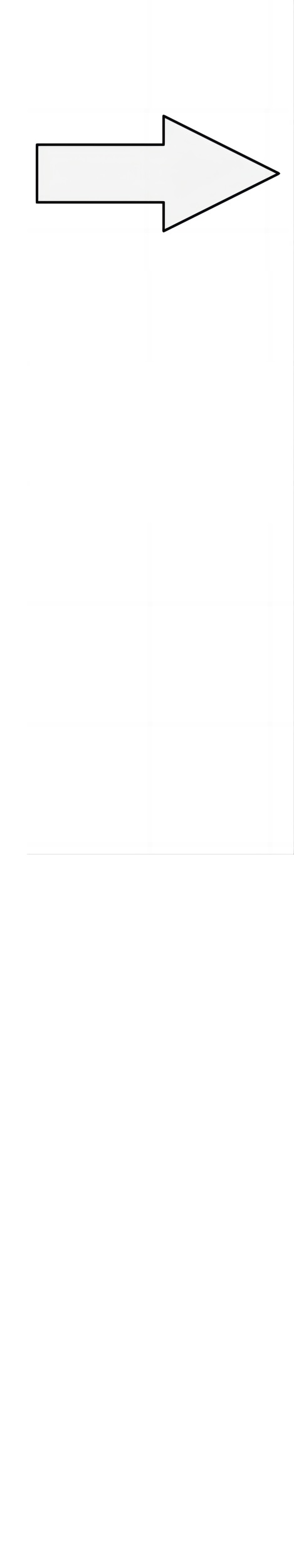}}
\subfigure[]{\includegraphics[width=7.7cm]{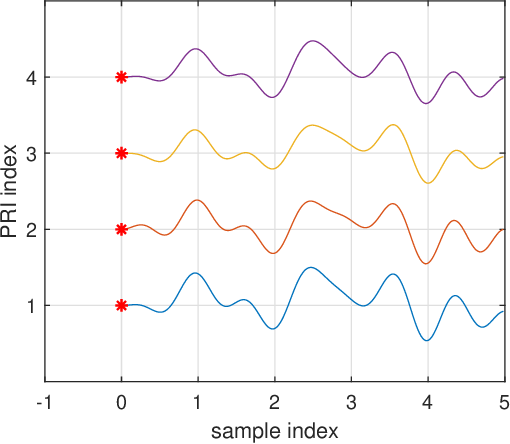}}\hspace{0.6cm}
\caption{Illustration of mapping 1-D raw data to 2-D raw data, $\mathbf{Y} = {\cal M}_{\rm 1D \rightarrow 2-D}(\mathbf{y}, P)$,  given a sample PRI $P$. This mapping is denoted as a parametric model ${\cal M}_{\rm 1D \rightarrow 2-D}(\mathbf{y}, P)$. (a) the 1-D signal, (b) form the 2-D data using the integral part of $P$, (c) align the start point of each PRI observation (this step requires interpolation and can be performed fast in the frequency domain). The vertical lines denote the sampling grids.}\label{fig:1d-2d-align}
\end{figure*}

First, we discuss the 2-D data formation model when the source PRI is known. The sample PRI is a parameter of this model, and we can handle the estimation of the PRI as a model parameter inversion problem.

The time axis $t'$ of the 1-D observed signal is related to the 2-D time axis of the 2-D signal as follows
\begin{equation}\label{}
t' = t + \tau = lT + \tau
\end{equation}
Here, the slow time is expressed as $t = lT$ where $l$ denotes the $l$-th pulse observation interval, considering that the slow time is discrete. The start time instance of the $l$-th pulse observation interval is $lT$. Accordingly, the 2-D signal can be expressed using the 1-D signal as follows
\begin{equation}\label{}
Y(t = lT, \tau) = y(lT + (0:lT)), \quad l = 1, \cdots, L
\end{equation}
The above model is expressed using continuous signals, but the sampled data is discrete. Thus, we need to construct a discrete 2-D data formation model.

The sample PRI can be expressed in the discrete domain as
\begin{equation}\label{}
P = T f_{\rm s} = P_{\rm int} + P_{\rm dec}
\end{equation}
where $P_{\rm int}$ is the integral part and $P_{\rm dec}$ is the decimal part. The start point of the $l$-th observation interval can be expressed as
\begin{equation}\label{}
\begin{split}
(l-1)P &= \text{int}((l-1)P) + \text{dec}((l-1)P)\\
         & = N_{l,\rm int} + N_{l,\rm dec}
\end{split}
\end{equation}
This point generally may not lie on the sampling grid, and the derivation of this point to the nearby sampling grid varies across different $l$. We must align the start point in the 2-D data matrix to preserve coherent information across different observation intervals.

The alignment of the $l$-th echo can be realized by shifting the data slice by a decimal sampling interval $N_{l,dec}$ in the discrete domain. This process can be expressed as
\begin{equation}\label{}
Y(l,:) = \text{shift}(y(N_{l,\rm int} + (0:P_{int})), N_{l,\rm dec})
\end{equation}
for $l = 1, \cdots, L$.
The shifting operation can be performed using FFT
\begin{equation}\label{}
\begin{split}
&\text{shift}(y(N_{l,\rm int} + (0:P_{\rm int} - 1)), N_{l,\rm dec}) \\
&= {\cal F}^{-1}({\cal F}(y(N_{l,\rm int} + (0:P_{\rm int} - 1))) W^{(0:P_{int} - 1)N_{l,\rm dec}})
\end{split}
\end{equation}
where $W = \exp(-j2\pi/P_{\rm int})$.
For the convenience of discussion, we use a mapping, $\mathbf{Y} = {\cal M}_{\rm 1D \rightarrow 2-D}(\mathbf{y}, P)$, to denote the above 2-D data matrix formation process from 1-D data given a known sample PRI $P$.

\subsection{Coarse Estimation of Sample PRI via Approximated Periodicity of 1-D Raw Data's Amplitude}

\begin{figure}
\centering
\subfigure[]{\includegraphics[width=7cm]{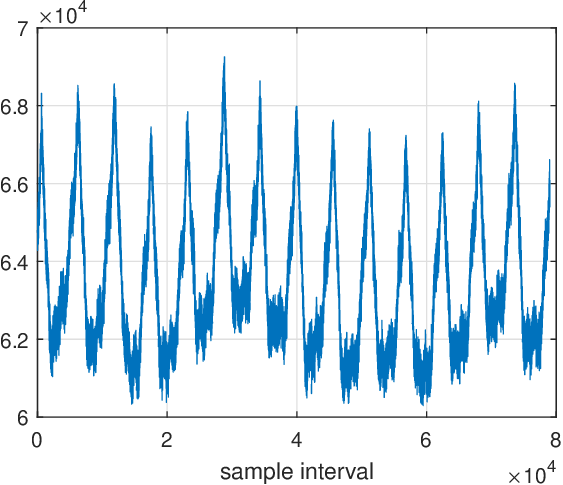}}
\subfigure[]{\includegraphics[width=7cm]{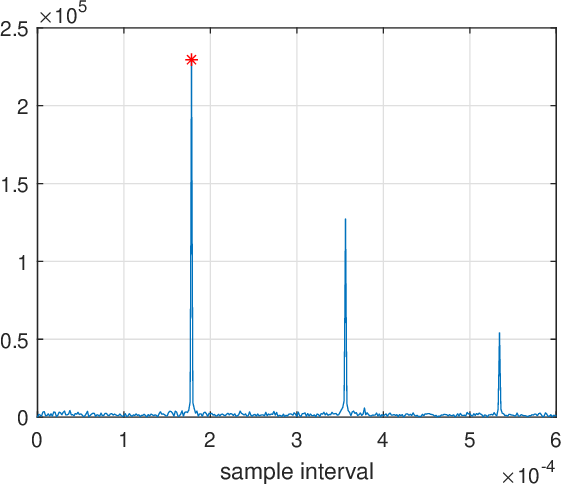}}
\caption{Examples of (a) cross correlating profile illustrating the approximated periodicity of  the 1-D data's amplitude, and (b) the coarse estimation of sample PRI from the spectrum of the 1-D data's amplitude. ERS raw data \url{E2_16385_STD_L0_F370} is used to obtain the results.}\label{fig:coarseT}
\end{figure}

The amplitude signals of different observation intervals are similar, so the full 1-D signal can be approximated as a periodic one, with $P$ being the period. This property can be expressed as
\begin{equation}
|Y(\eta, \tau)| = |Y(\eta + PRI, \tau)| \Rightarrow |y(t')| \approx |y(t' + PRI)|
\end{equation}
This periodicity can be illustrated by cross correlating a slice of the data with the rest data, which is expect to have periodical peaks. An example of such behavior is shown in Fig. \ref{fig:coarseT}(a) using ERS data.
Due to the periodicity, the spectrum of the 1-D amplitude will have a peak at $1/P$ on the normalized frequency axis in $[0, 1]$.
Therefore, we can estimate the period by finding the non-zero frequency peak location $f_{\rm peak}$ of the FFT spectrum and take its inverse as the PRI estimate $\hat{P}_0 = 1/f_{\rm peak}$. An example of such estimation is provided in Fig. \ref{fig:coarseT}, where the $\hat{P}_0 = 5614.03$. However, such estimation is coarse and requires further fine estimation.

\subsection{Fine Estimation of Sample PRI via PCM}

We propose using PCM for fine estimation, which can be expressed as an optimization problem
\begin{equation}
\begin{array}{lll}
\hat{P}=
& \mathop{\rm argmax}\limits_{P} & \|\mathbf{Y}_{\rm sub, pc}\|_F^2  \\
& \hbox{subject to} & \mathbf{Y}_{\rm sub, pc} = {\cal P}\left(\mathbf{Y}_{\rm sub},D=1\right)\\
&                              & \mathbf{Y}_{\rm sub} = {\cal M}(\mathbf{y}_{\rm sub}, P)
\end{array}
\end{equation}
where $\mathbf{y}_{\rm sub}$ is a selected subset (e.g., its first 100,000 samples) of the full 1-D raw data. This problem is equivalent to maximizing the largest eigenvalue $\rho_1$ of the data covariance matrix $\mathbf{Y}_{\rm sub} \mathbf{Y}_{\rm sub}^{\rm H}$.
We solve this problem via Q stages of searching. For the $q$-th stage, the searching interval is $\hat{P}_{q-1} + [-W_{q-1}, W_{q-1}]$ using searching grid spacing $\Delta P_q < 1$. The searching result is denoted as $\hat{P}_{q}$, and it is used as the center point in the next stage searching. In the first stage, we use $\hat{P}_0$ obtained in the first step exploiting the amplitude's periodicity.

\begin{figure}
\centering
\subfigure[]{\includegraphics[width=7cm]{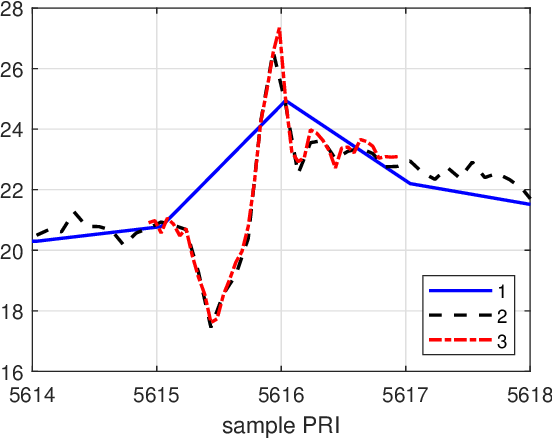}}
\caption{Example of fine sample PRI estimation. The peak points in the 3 stage searching are $5616.04$, $5615.84$ and $5615.98$, respectively. The ground truth is $P = 5616$. ERS raw data \url{E2_16385_STD_L0_F370} is used to obtain the results.}\label{fig:eig_PRI}
\end{figure}

In summary, we estimate the sample PRI via a two-step process based on the 1-D data's approximated periodicity model and principal component maximization:
\begin{enumerate}
\item Estimate the coarse sample PRI using the FFT peak of the amplitude.
\item Perform fine sample PRI estimation using PCM.
\end{enumerate}
Fig. \ref{fig:eig_PRI} illustrates the fine estimation of sample PRI using three stages searching starting with a  coarse estimate $\hat{P}_0=5614.03$. The estimation results in the 3 stage searching  are $5616.04$, $5615.84$ and $5615.98$, respectively, which gradually achieve the true sample PRI value $P = 5616$.

\section{Phase Error Analysis and Correction}\label{sec:err-correct}

In the above content, we used an approximated shift-invariant model for image formation using the reference echo as the matched filter.  This approximation will leads to phase errors in the azimuth direction and finally cause azimuth defocusing. In this section, we detail the phase error analysis and the method for phase error correction.

\subsection{Phase Error of the Approximated Shift-Invariance Model}

In the image formation model discussed in Section III, the following approximation is used
\begin{equation}
R(\eta,p+\Delta p) \approx R + \Delta R + \frac{(v\eta-x-\Delta x)^2}{2R}
\end{equation}
This approximation allows us to obtain a shift-invariant model for image formation using the reference echo as the matched filter
\begin{equation}
Y_{p+\Delta p} \approx Y_p(\eta-\Delta x / v, \tau-2\Delta R / c)
\end{equation}
Using this approximation, the azimuth signal of $p+\Delta p$ can be written as
\begin{equation}\label{}
Y_{{\rm a},p+\Delta p}(\eta) = \exp\left(-j\frac{4\pi}{\lambda}\frac{(v\eta-x-\Delta x)^2}{2(R+\Delta R)}\right)
\end{equation}
This expression can be approximated using a shifted version of the azimuth signal of the reference point $p$ as follows
\begin{equation}\label{}
Y_{{\rm a},p}(\eta-\eta_0) = \exp\left(-j\frac{4\pi}{\lambda}\frac{(v\eta-x)^2}{2R}\right)
\end{equation}

However, the approximation of the range history introduces a quadratic phase error, which can be expressed as follows
\begin{equation}
\begin{split}
\left|\frac{2\pi(v\eta-x)^2}{\lambda R_0} -
\frac{2\pi(v\eta-x)^2}{\lambda(R_0+\Delta R)} \right|
=
\frac{2\pi(v\eta-x)^2\left|\Delta R\right|}{\lambda R_0(R_0+\Delta R)}
\end{split}
\end{equation}

The phase error causes azimuth defocusing to various degrees, depending on the range of the reference echo and the distance between the range cell and the reference point. The defocusing can be analyzed in the azimuth frequency domain using azimuth LFM signal models. Specifically, the azimuth FM rates of $Y_{{\rm a},p+\Delta p}(\eta)$ and $Y_{{\rm a},p}(\eta)$ are
\begin{subequations}\label{}
\begin{align}
K_{{\rm a},p+\Delta p} &= -\frac{2v^2}{\lambda(R_0+\Delta R)} \\
K_{{\rm a},p} &= -\frac{2v^2}{\lambda R_0}
\end{align}
\end{subequations}
respectively. Using the echo of $p$ for matched filtering results in a residual azimuth FM rate
\begin{equation}\label{}
K_{\rm a,residual} = \left| \frac{1}{\frac{1}{K_{{\rm a},p}} - \frac{1}{K_{{\rm a},p+\Delta p}}} \right|= \frac{2v^2}{\lambda |\Delta R|}
\end{equation}

The azimuth signal is not focused on a single point, but over an azimuth time span expressed as follows
\begin{equation}\label{}
\zeta_{\Delta R} = \frac{B_{\rm a}}{K_{\rm a,residual}} + \frac{1}{B_{\rm a}}
\end{equation}
where $B_{\rm a}$ is the azimuth bandwidth, and $\frac{1}{B_{\rm a}}$ accounts for the azimuth sidelobes. Assuming the SAR antenna has an azimuth length $L_{\rm a}$ and its beam pointing is fixed, the azimuth bandwidth can be expressed as
\begin{equation}\label{}
B_{\rm a} = \frac{2v}{L_{\rm a}} {\rm sinc}\left(\frac{\lambda}{2L_{\rm a}}\right)
\end{equation}
Using this expression of $B_{\rm a}$ along with the definition of $K_{\rm a,residual}$, we can express the azimuth time span as
\begin{equation}\label{}
\zeta_{\Delta R} = \frac{\lambda |\Delta R|}{vL_{\rm a}} {\rm sinc}\left(\frac{\lambda}{2L_{\rm a}}\right) + \frac{1}{B_{\rm a}}
\end{equation}
The number of azimuth resolution elements in this azimuth time span can be expressed as
\begin{equation}\label{}
\zeta_{\Delta R} \cdot B_{\rm a} = \frac{2\lambda |\Delta R|}{L_{\rm a}^2} {\rm sinc}^2\left(\frac{\lambda}{2L_{\rm a}}\right) + 1
\end{equation}
From the error expression, we can see that this error is large for small $L_{\rm a}$ (i.e., when the scene is at a far range). Therefore, for airborne SAR, this error should be compensated to improve image quality.

\subsection{Correction of the Model Phase Error via Range-Varying Azimuth FM Rate Estimation}

The quadratic phase error can be compensated using the following steps. First, perform range matched filtering using the range signal estimated from the reference echo. Second, estimate the azimuth reference signal $u(n,m)=\exp(j\pi \tilde{K}_{\rm a}(m)n^2)$ for each range cell. Here $\tilde{K}_{\rm a}$ is the azimuth FM rate normalized by the squared PRF. Third, perform azimuth compression via matched filtering for each range cell using the estimated azimuth reference signal.

The estimation of the azimuth reference signal for each range cell can be performed in two steps. First, divide the range-compressed signal into $M_{\rm blk}$ blocks with center range cells $m_l$, $l=1,\cdots,M_{\rm blk}$. Second, estimate the azimuth FM rate $\tilde{K}_{\rm a}(m_l)$ for each block via Maximum Likelihood Estimation (MLE). Finally, perform FM rate fitting on $\{(m_l, \tilde{K}_{\rm a}(m_l)) | l=1,\cdots,M_{\rm blk}\}$ to obtain the FM rate $\tilde{K}_{\rm a}(m)$ for each range cell $m=1,\cdots,M_{\rm  blk}$.

\section{Experiment} \label{sec:exp}

We show several experimental results to validate the effectiveness of proposed method in recovering images from SAR raw data. The experiments involve processing ERS data,  RADARSAT data, and airborne SAR data using the described algorithm using given block sizes (for the convenience of discussion, we refer to it as method 1 in the following). For comparison, we also generate SAR images using another two methods:  the non block-segmentation version of the proposed method, i.e., the full data is treated as a single block (we refer to it as method 2 in the following),  and  the chirp scaling algorithm (we refer to it as method 3 in the following). All the resulting images are multi-looked for visualization, and detailed image patches are shown without multi-looking.

%

\begin{figure}
  \centering
  \subfigure[]{\includegraphics[width=8.8cm]{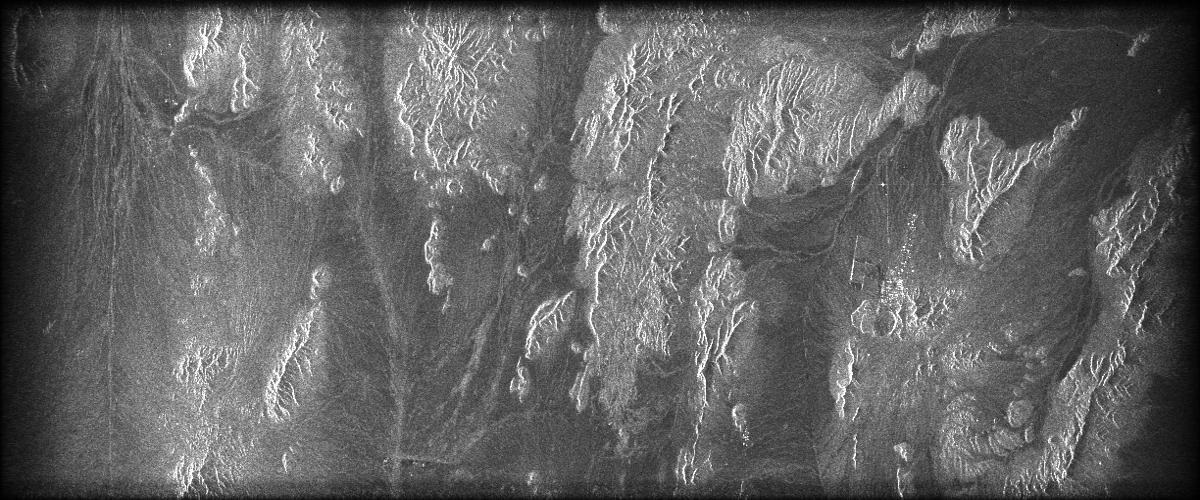}}
  \subfigure[]{\includegraphics[width=8.8cm]{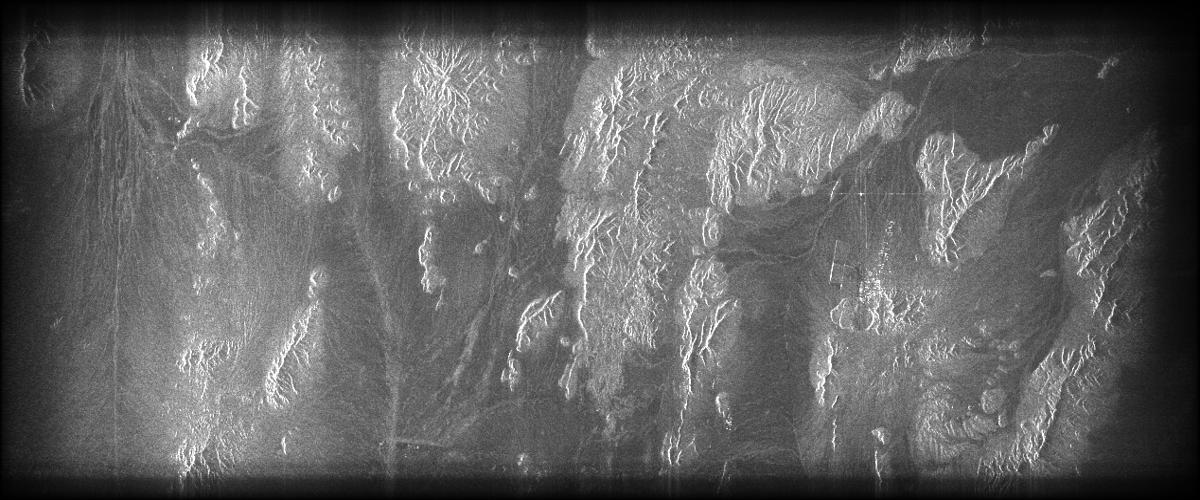}}
  \subfigure[]{\includegraphics[width=8.8cm]{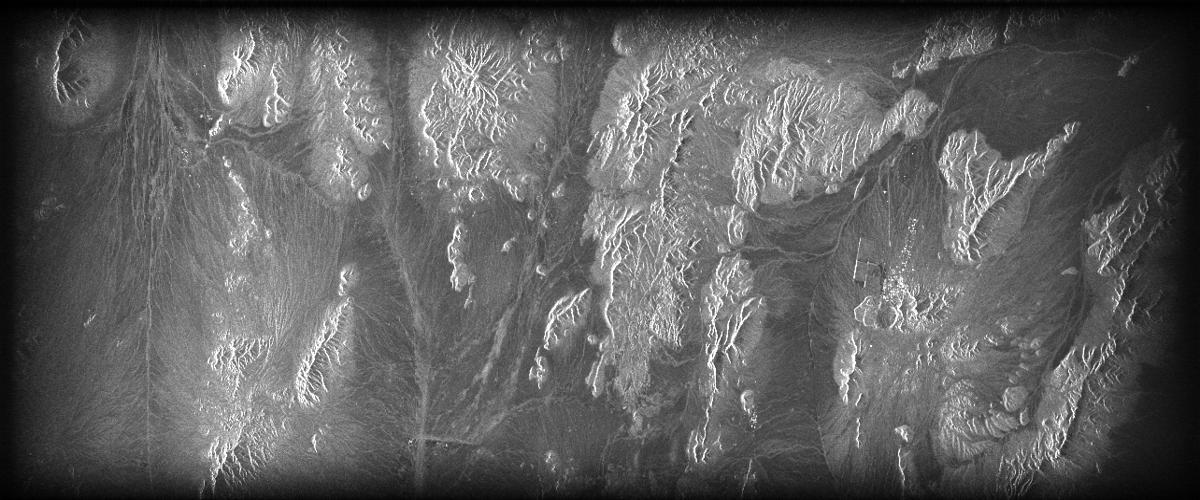}}
    \subfigure[]{\includegraphics[width=2.85cm,height=2.85cm]{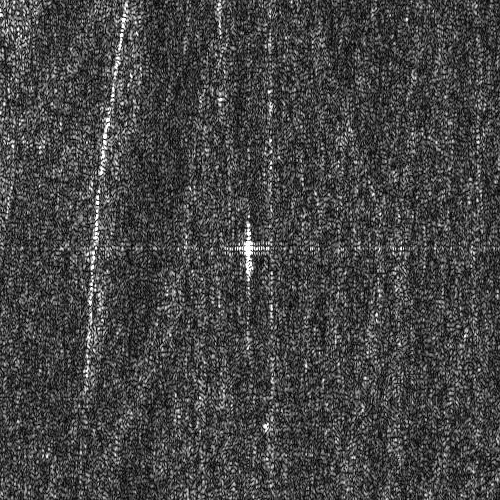}}
  \subfigure[]{\includegraphics[width=2.85cm,height=2.85cm]{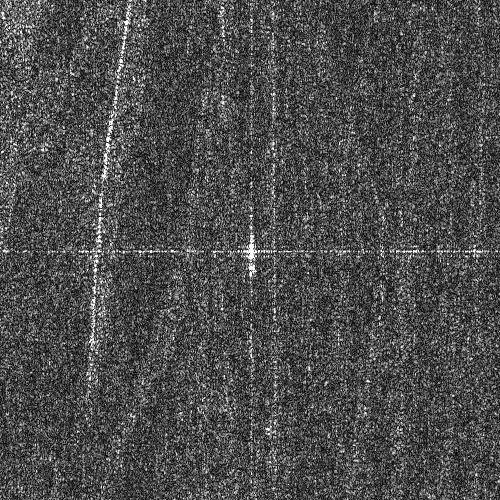}}
  \subfigure[]{\includegraphics[width=2.85cm,height=2.85cm]{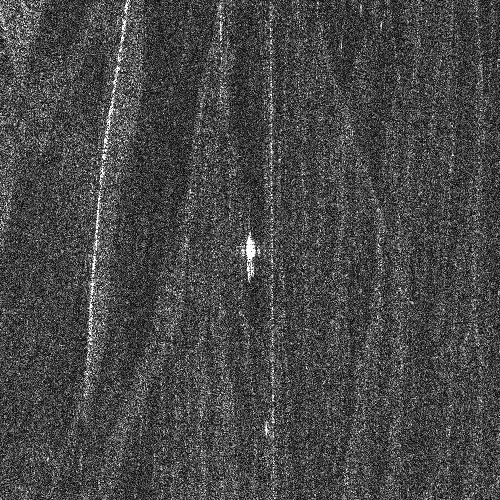}}
  \caption{The generated images in experiment \#1. (a) by method 1, (b) by method 2, (c) by method 3. (d)-(f) are detailed patches selected from (a)-(c), respectively.}\label{fig:e1}
\end{figure}

\begin{figure}
  \centering
  \subfigure[]{\includegraphics[width=8.8cm]{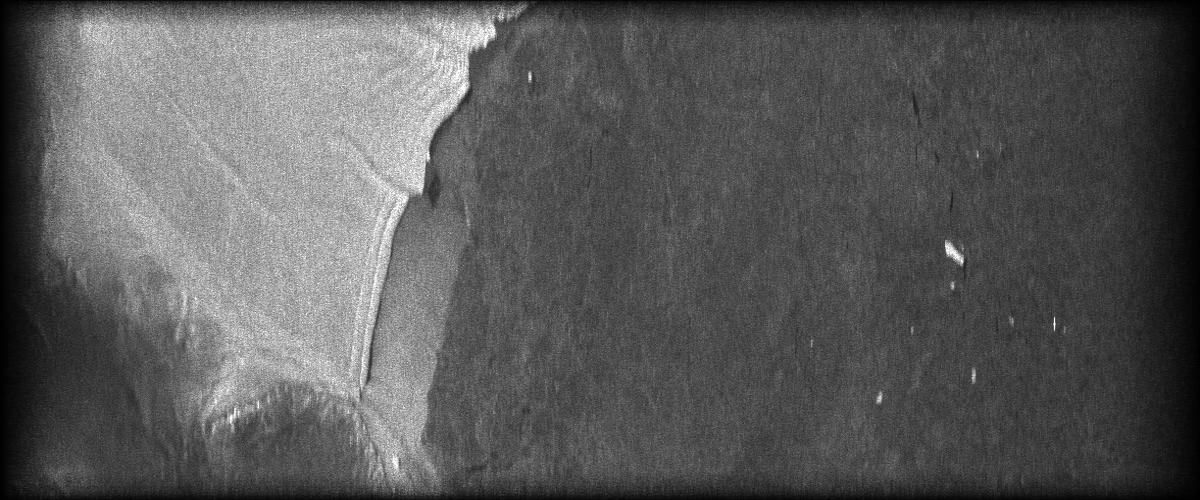}}
  \subfigure[]{\includegraphics[width=8.8cm]{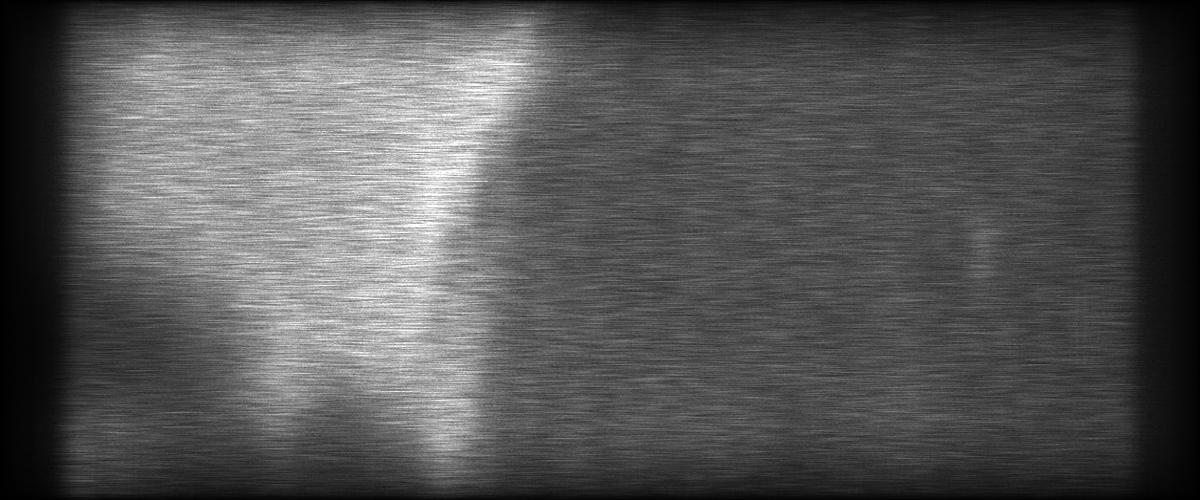}}
  \subfigure[]{\includegraphics[width=8.8cm]{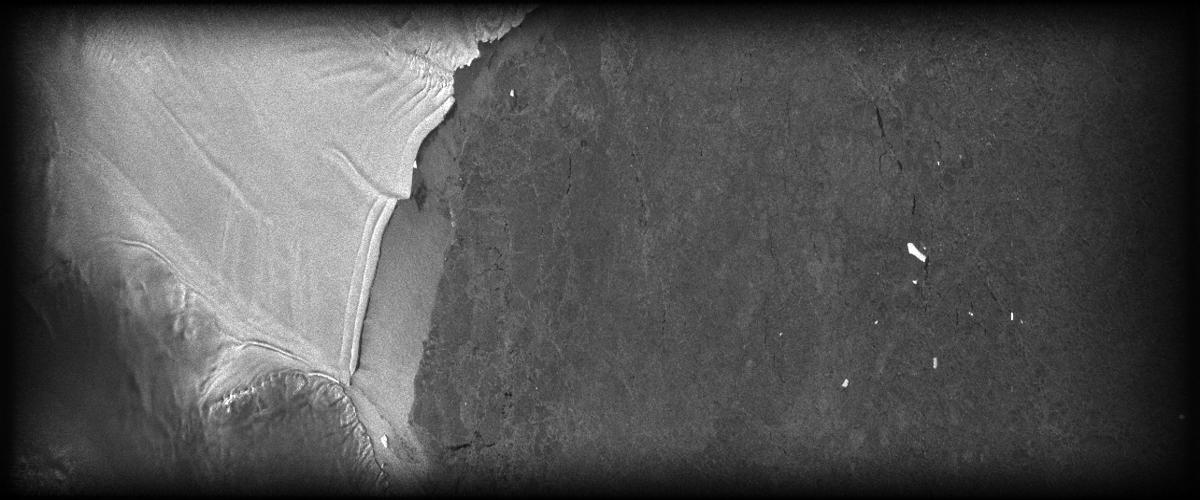}}
    \subfigure[]{\includegraphics[width=2.85cm,height=2.85cm]{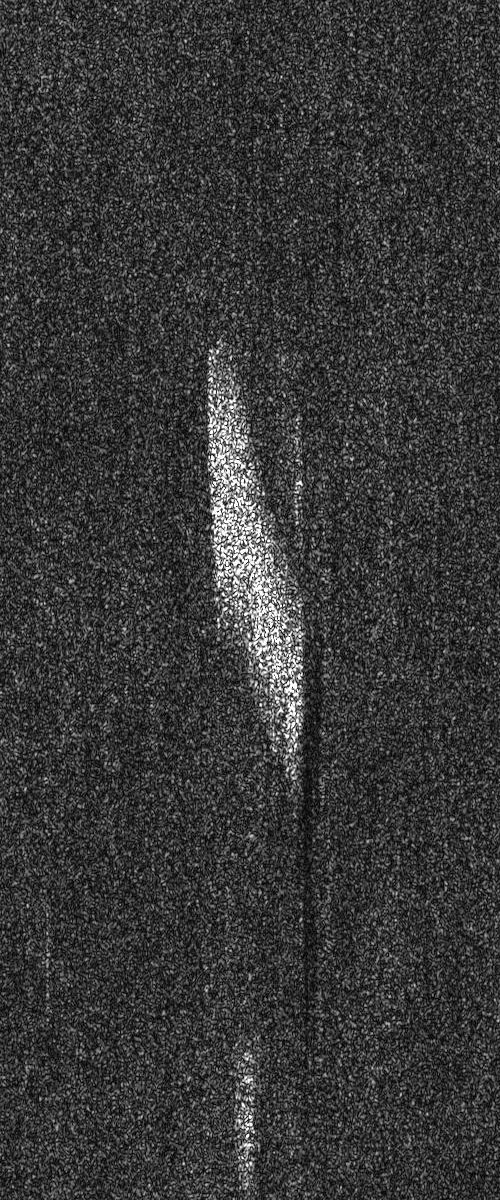}}
  \subfigure[]{\includegraphics[width=2.85cm,height=2.85cm]{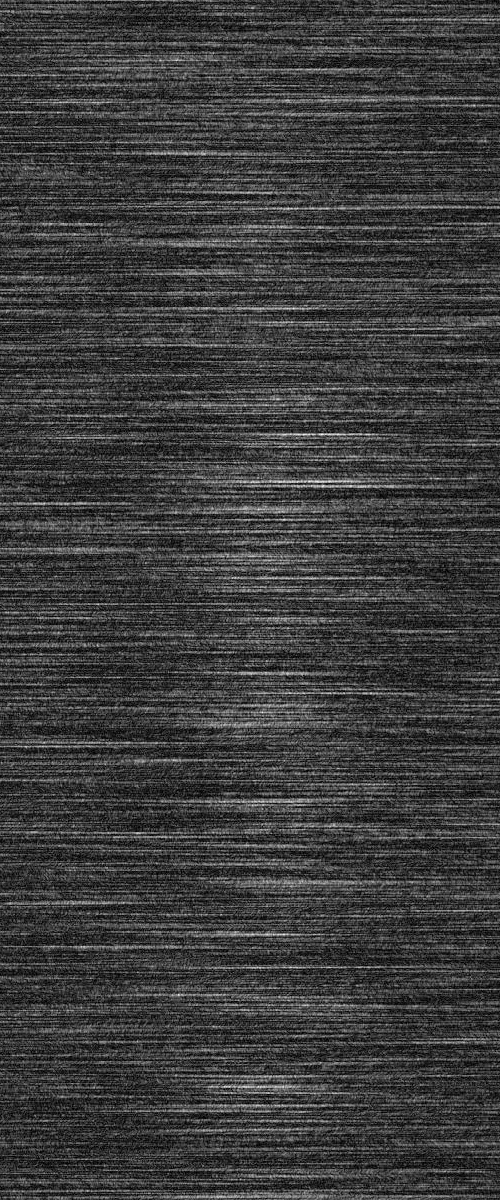}}
  \subfigure[]{\includegraphics[width=2.85cm,height=2.85cm]{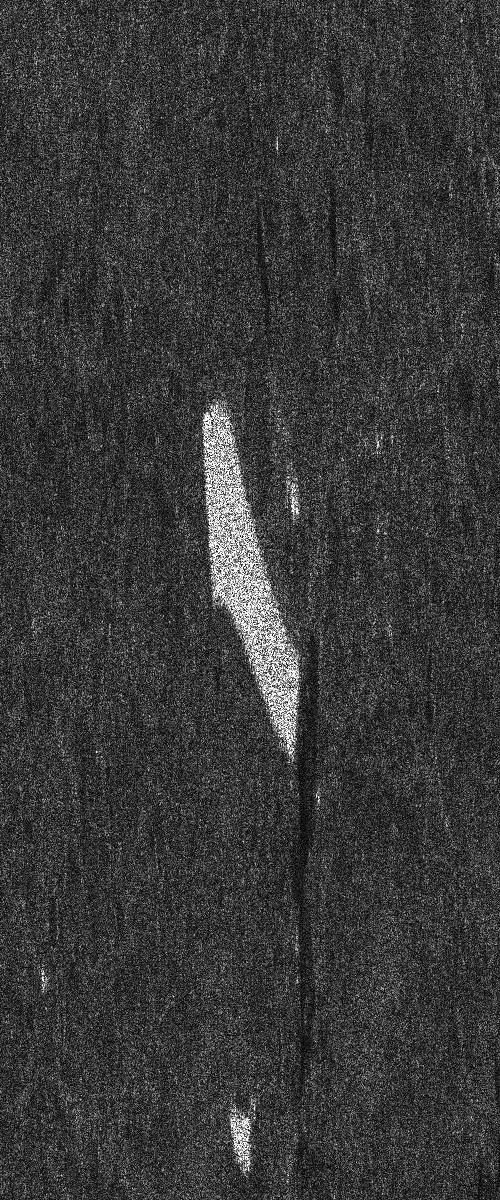}}
  \caption{The generated images in experiment \#2. (a) by method 1, (b) by method 2, (c) by method 3 (chirp scaling algorithm). (d)-(f) are detailed patches selected from (a)-(c), respectively.}\label{fig:e2}
\end{figure}

\subsection{Experiment \#1 using ERS Raw Data}

In this experiment we use ERS raw data to test the proposed method. The data id is \url{E2_16385_STD_L0_F370} accessed from the Alaska Satellite Facility \url{https://search.asf.alaska.edu/}. The block size used in our method is $500\times 500$. The generated images using the proposed method and the other two methods are provided in Fig. \ref{fig:e1}. Specifically, Fig.  \ref{fig:e1}(a) is the image generated by the proposed method using block size $500\times 500$, Fig.  \ref{fig:e1}(b) is the image generated by the non block-segmentation version of the proposed method, and Fig.  \ref{fig:e1}(c) is the image generated by the chirp scaling method. In this data, there exists a dominating point scatterer at the right-top area in the image domain. The proposed method with block segmentation can effectively estimate the point scatterer echo and generates a high-quality image. Because the point scatterer is very strong, the non block-segmentation version can also successfully reconstruct a high-quality image, which is similar to that obtained by the block-segmentation version. However, the images' quality is inferior to that obtained by the chirp scaling method. To clearly compare the images, we show the single-look patches around the strong point scatterer in Figs. \ref{fig:e1}(d)-(f), which are extracted from the three images.  The results show that the patches by methods 1 and 2 are blurred compared to that by the method 3 (chirp scaling algorithm).
\begin{figure}
  \centering
  \subfigure[]{\includegraphics[width=8.8cm]{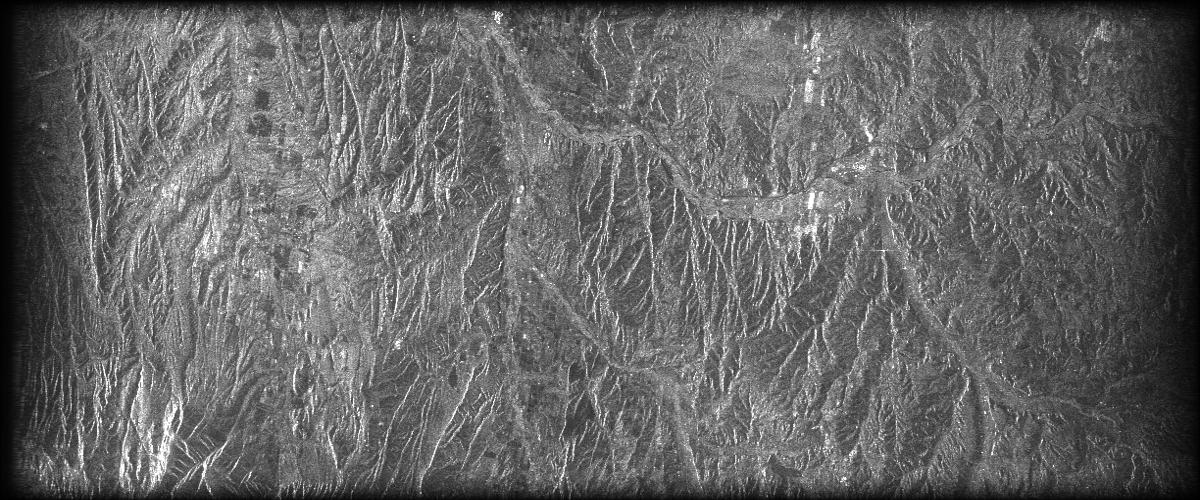}}
  \subfigure[]{\includegraphics[width=8.8cm]{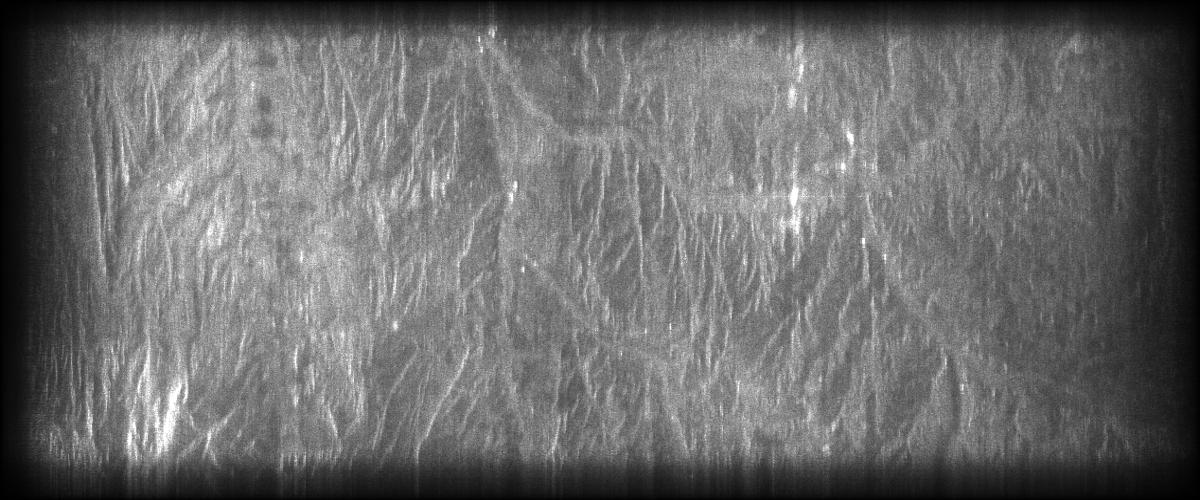}}
  \subfigure[]{\includegraphics[width=8.8cm]{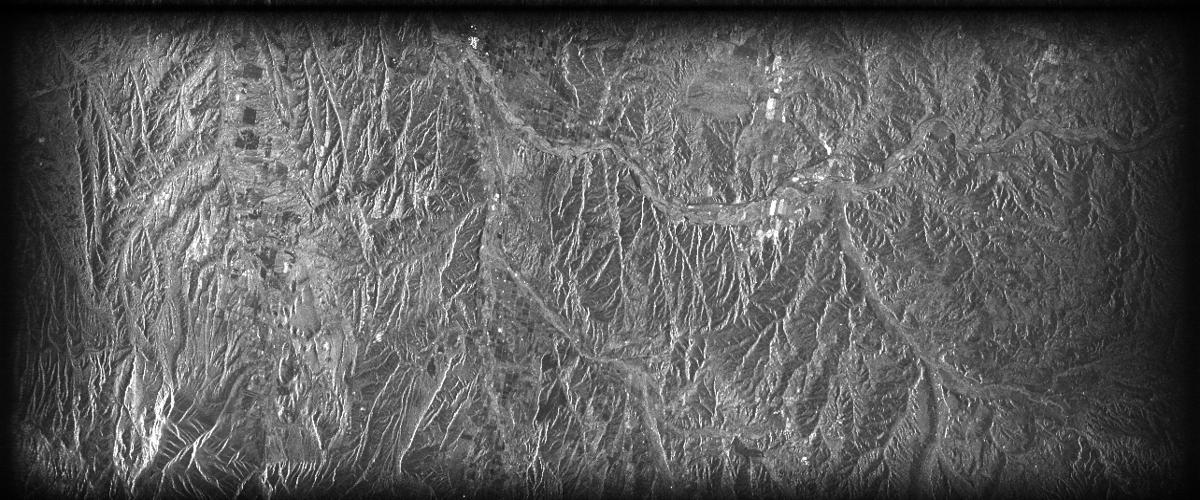}}
    \subfigure[]{\includegraphics[width=2.85cm,height=2.85cm]{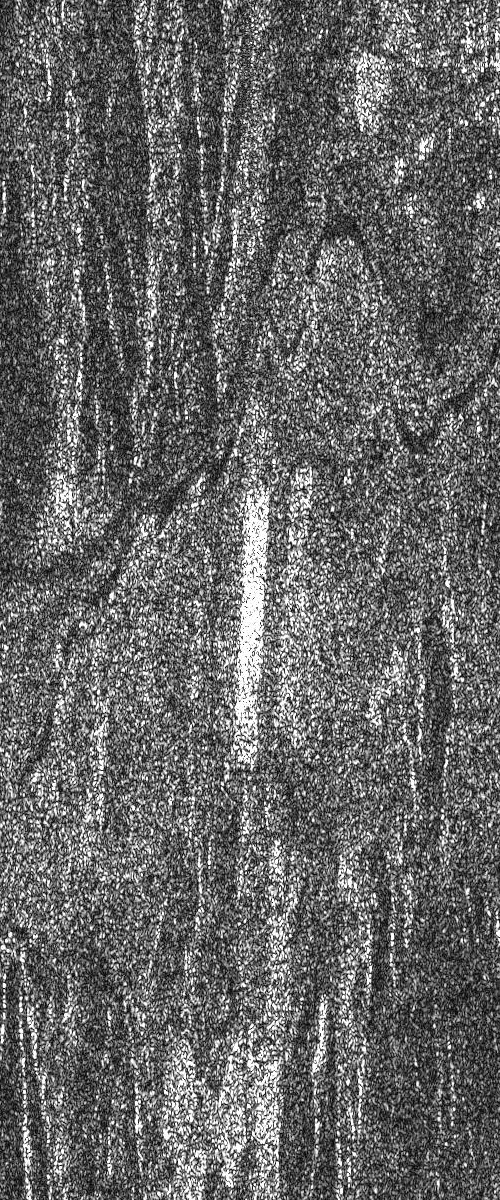}}
  \subfigure[]{\includegraphics[width=2.85cm,height=2.85cm]{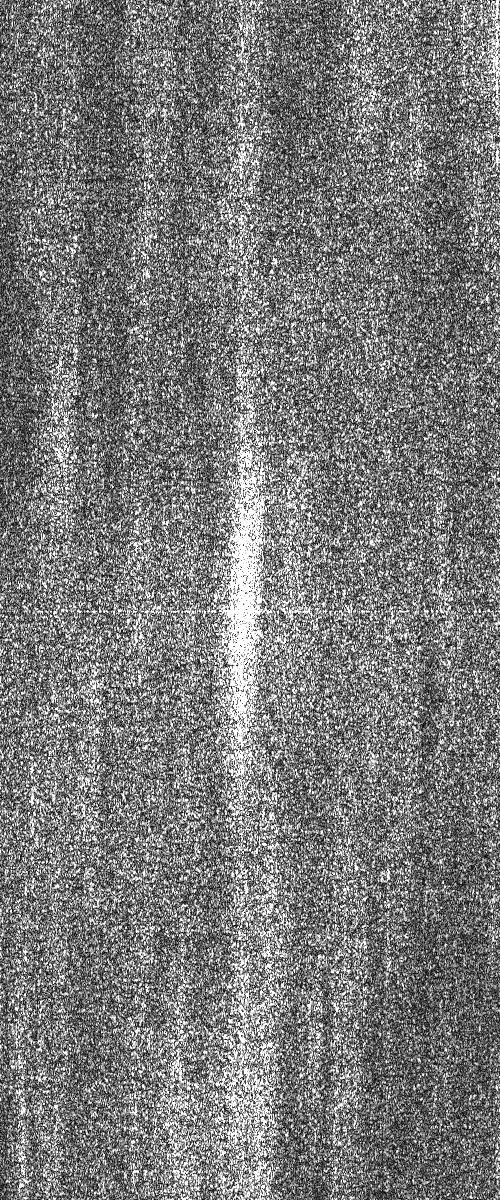}}
  \subfigure[]{\includegraphics[width=2.85cm,height=2.85cm]{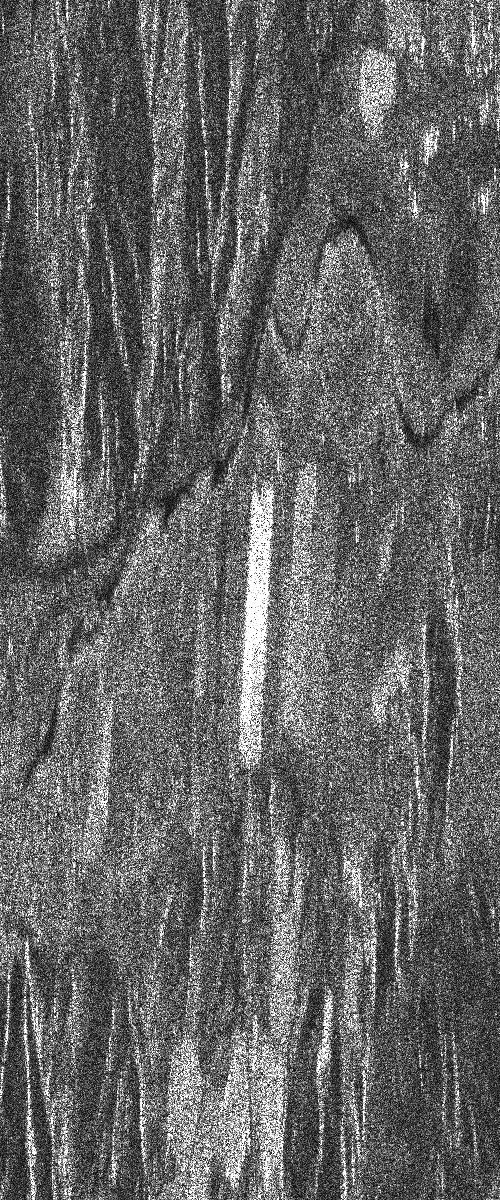}}
  \caption{The generated images in experiment \#3. (a) by method 1, (b) by method 2, (c) by chirp scaling algorithm. (c)-(d) are  and detailed patches selected from (a)-(c), respectively.}\label{fig:e3}
\end{figure}

\begin{figure}
  \centering
  \subfigure[]{\includegraphics[width=8.8cm]{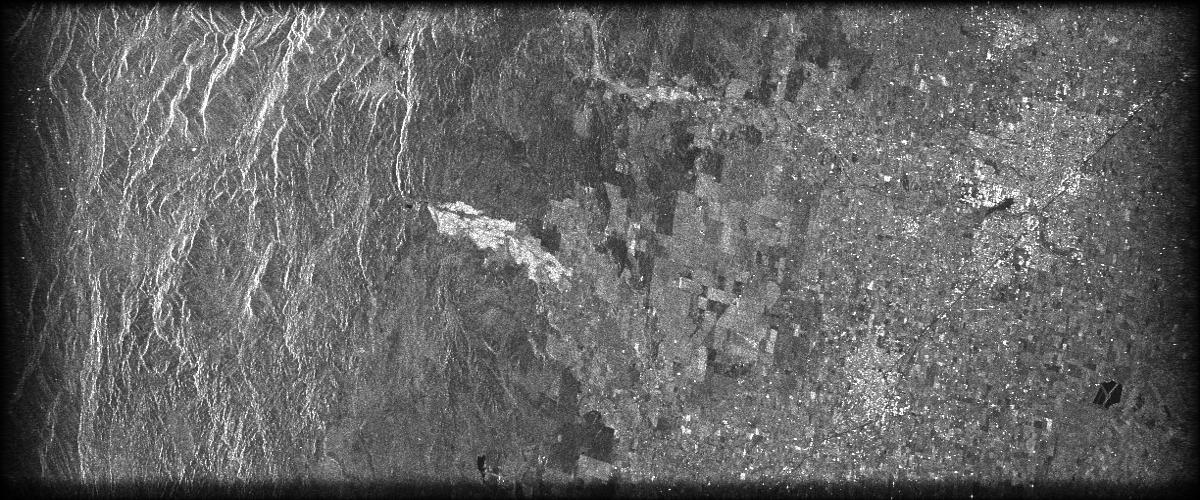}}
  \subfigure[]{\includegraphics[width=8.8cm]{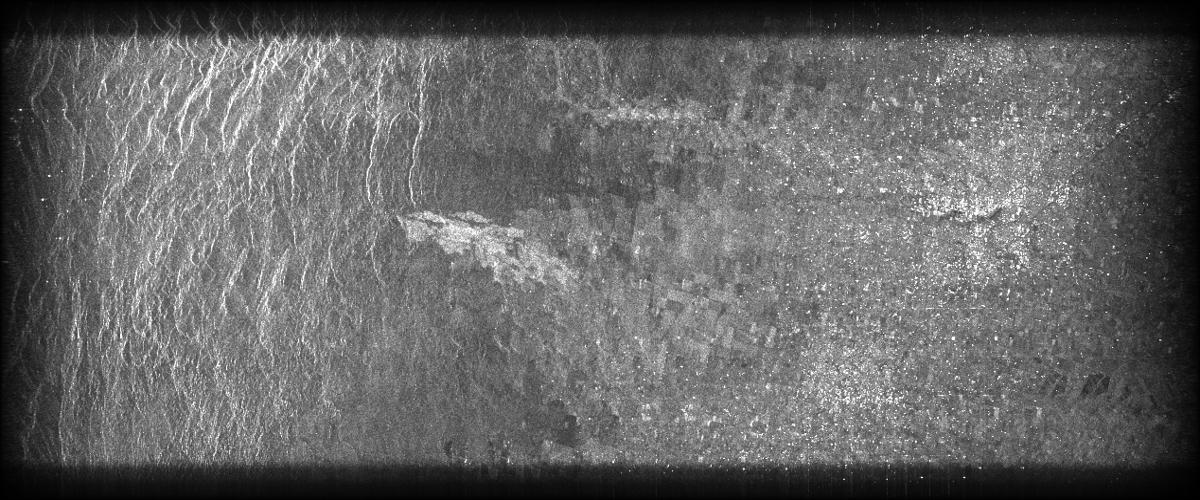}}
  \subfigure[]{\includegraphics[width=8.8cm]{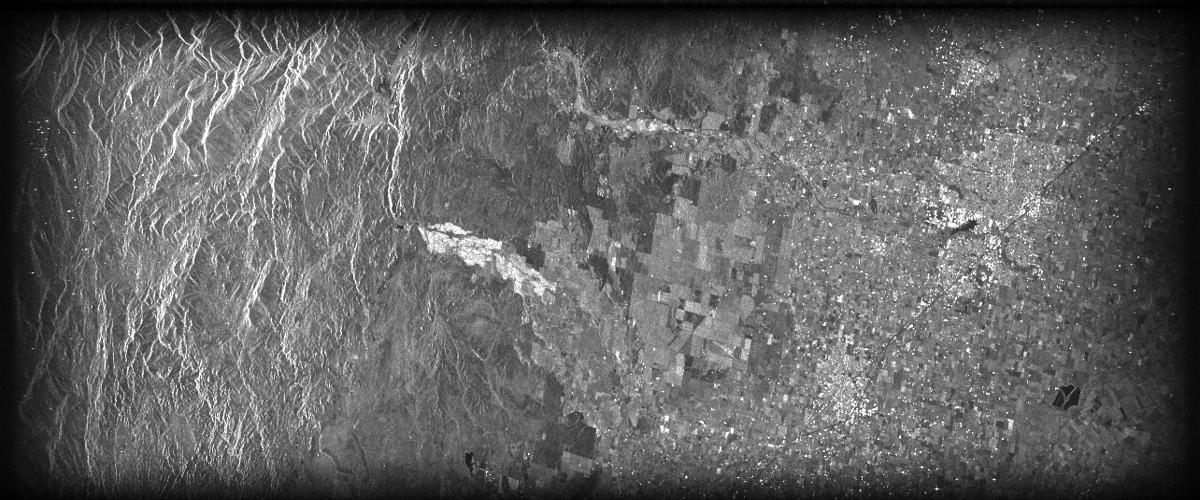}}
    \subfigure[]{\includegraphics[width=2.85cm,height=2.85cm]{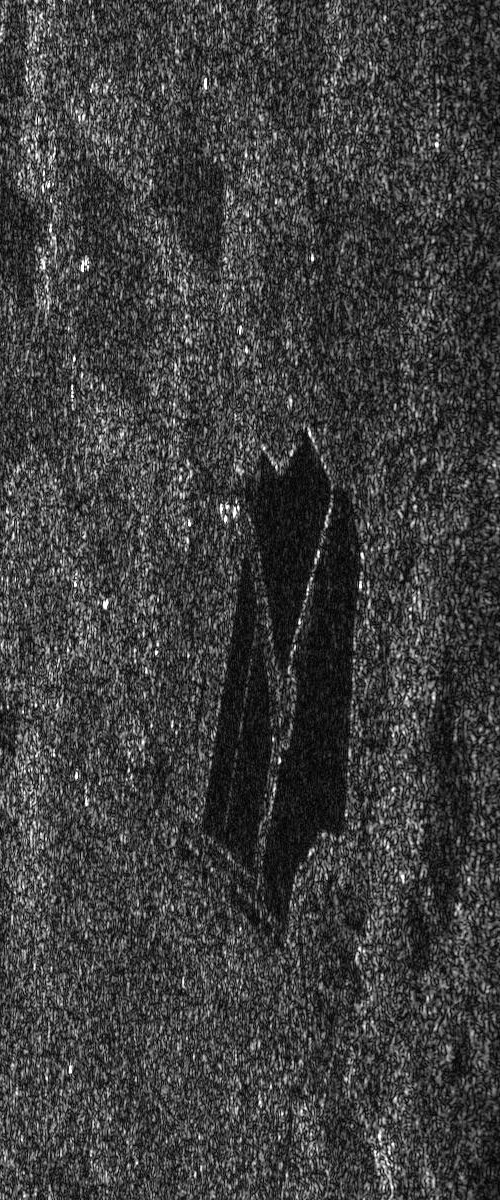}}
  \subfigure[]{\includegraphics[width=2.85cm,height=2.85cm]{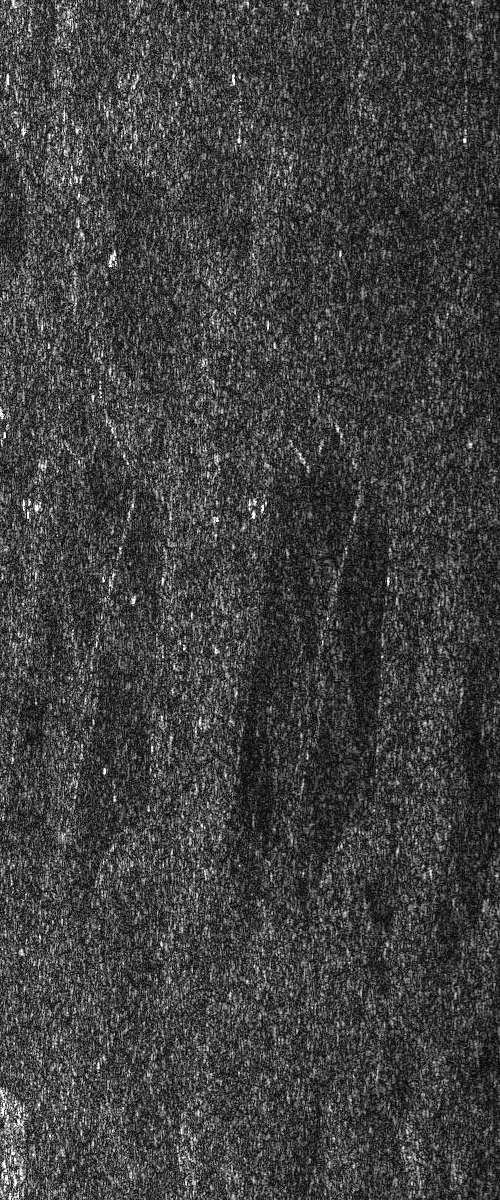}}
  \subfigure[]{\includegraphics[width=2.85cm,height=2.85cm]{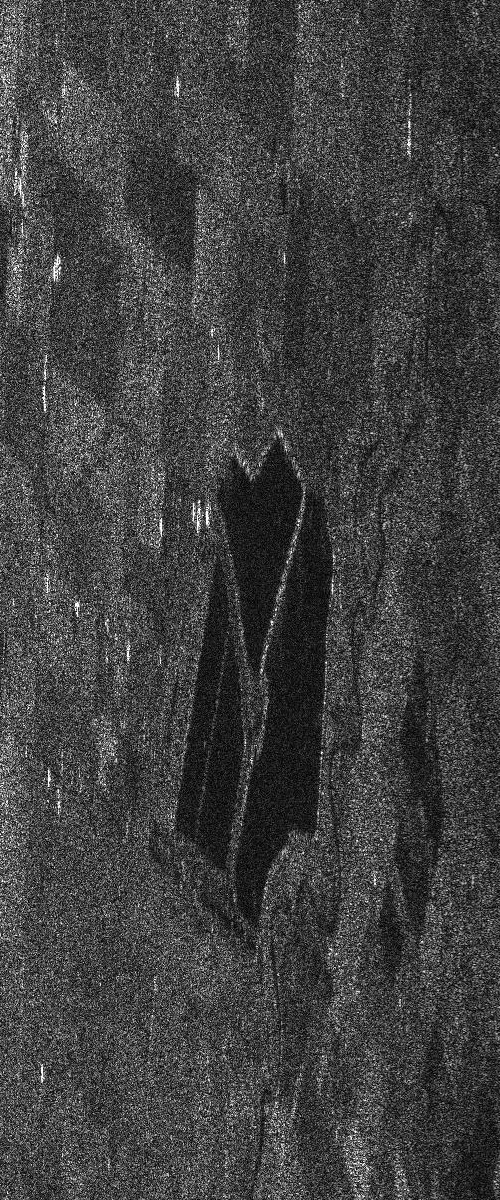}}
  \caption{The generated images in experiment \#4. (a) by method 1, (b) by method 2, (c) by method 3. (d)-(f) are detailed patches selected from (a)-(c), respectively.}\label{fig:e4}
\end{figure}

\begin{figure}
  \centering
  \subfigure[]{\includegraphics[width=8.8cm,height=3.67cm]{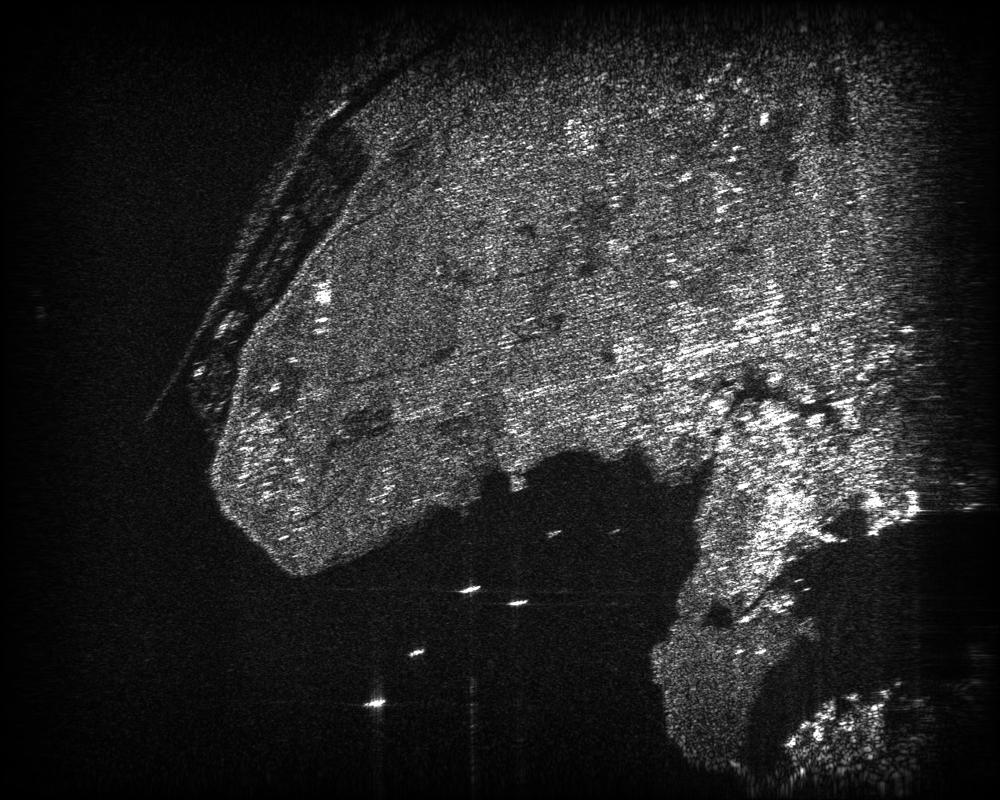}}
  \subfigure[]{\includegraphics[width=8.8cm,height=3.67cm]{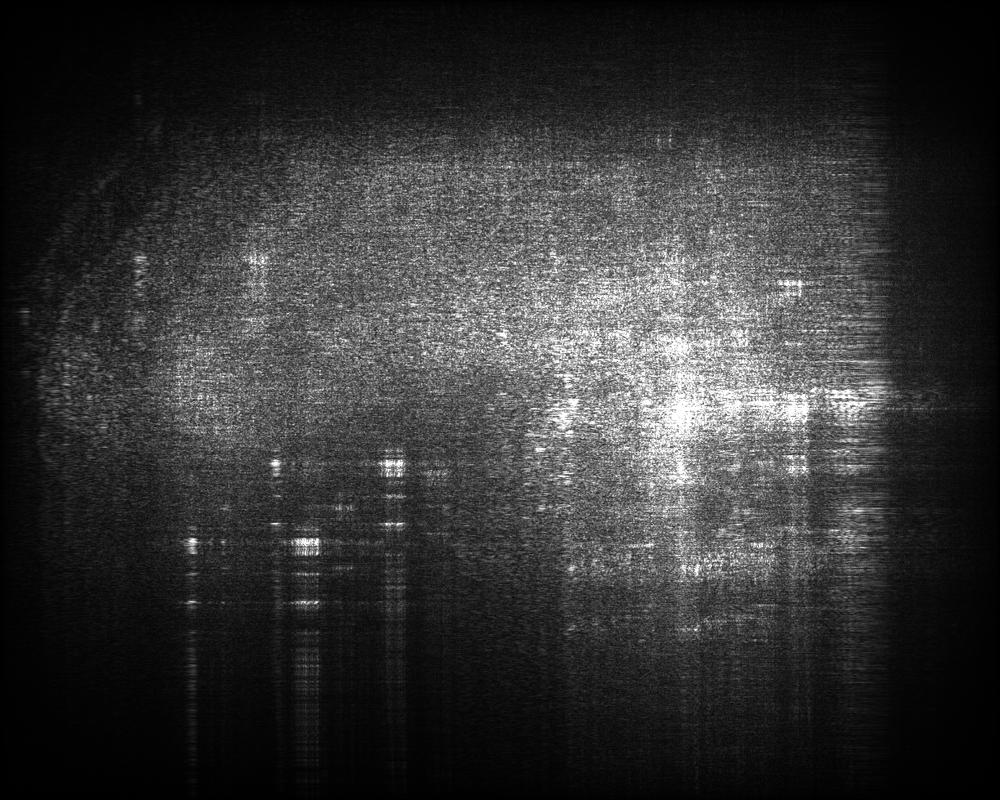}}
  \subfigure[]{\includegraphics[width=8.8cm,height=3.67cm]{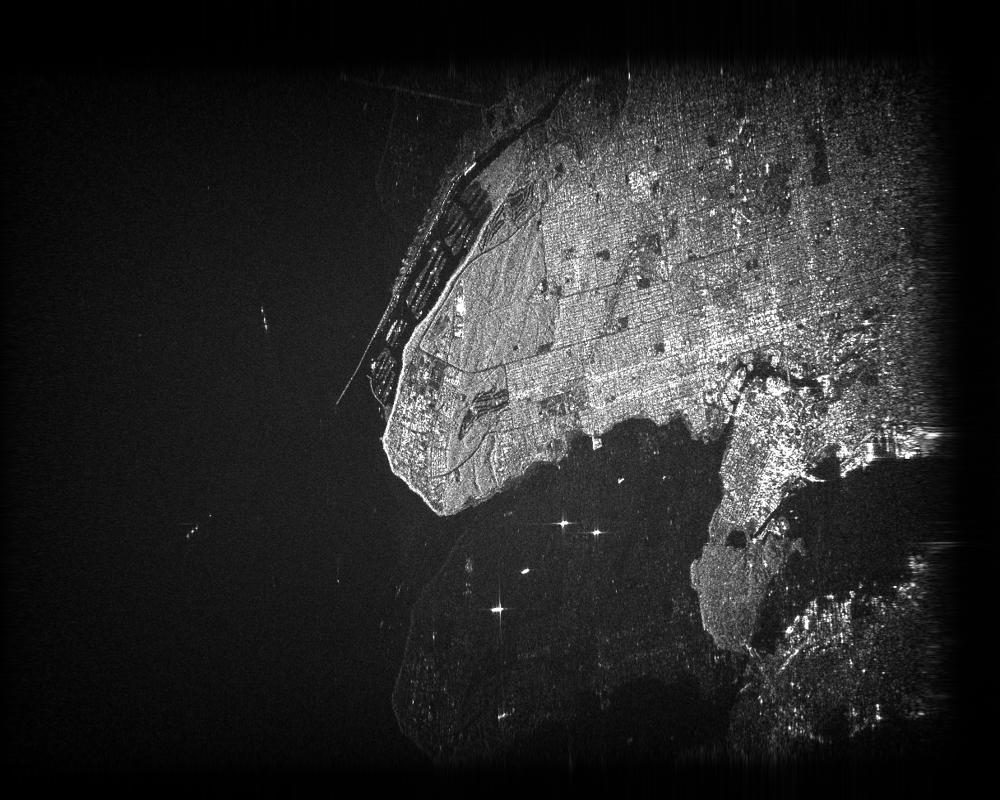}}
    \subfigure[]{\includegraphics[width=2.85cm,height=2.85cm]{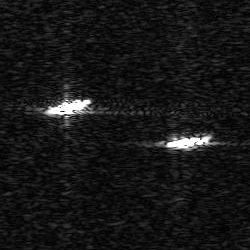}}
  \subfigure[]{\includegraphics[width=2.85cm,height=2.85cm]{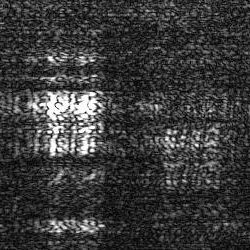}}
  \subfigure[]{\includegraphics[width=2.85cm,height=2.85cm]{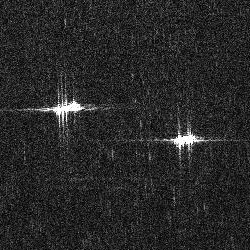}}
  \caption{The generated images in experiment \#5. (a) by method 1, (b) by method 2, (c) by method 3. (d)-(f) are detailed patches selected from (a)-(c), respectively.}\label{fig:e5}
\end{figure}

\begin{figure}
  \centering
  \subfigure[]{\includegraphics[width=8.8cm]{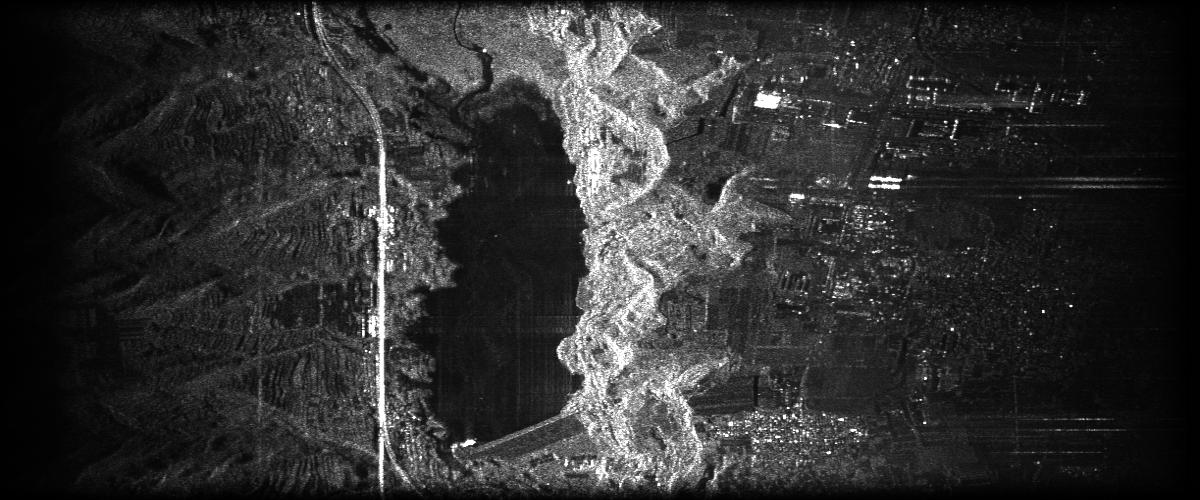}}
  \subfigure[]{\includegraphics[width=8.8cm]{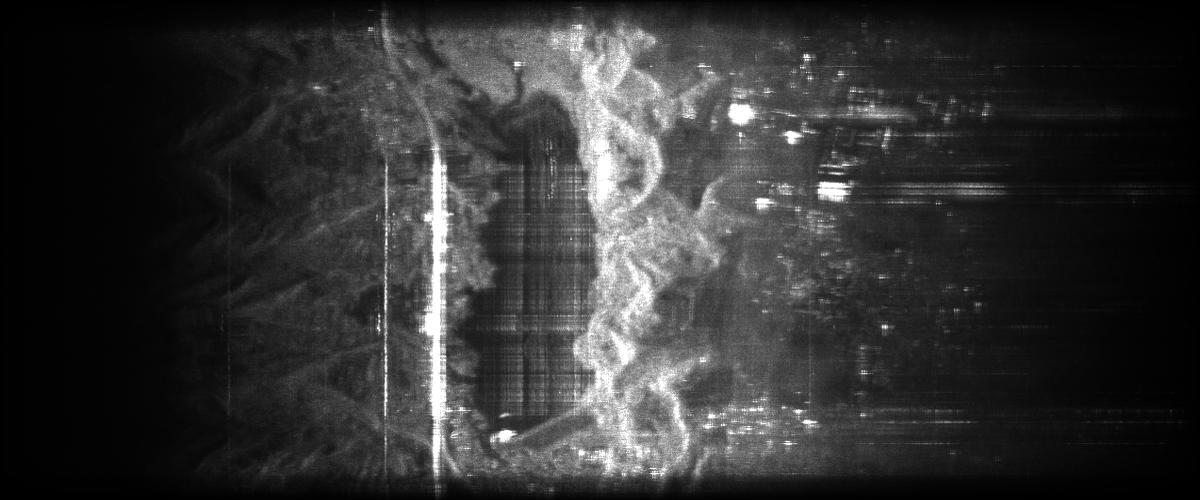}}
  \subfigure[]{\includegraphics[width=8.8cm]{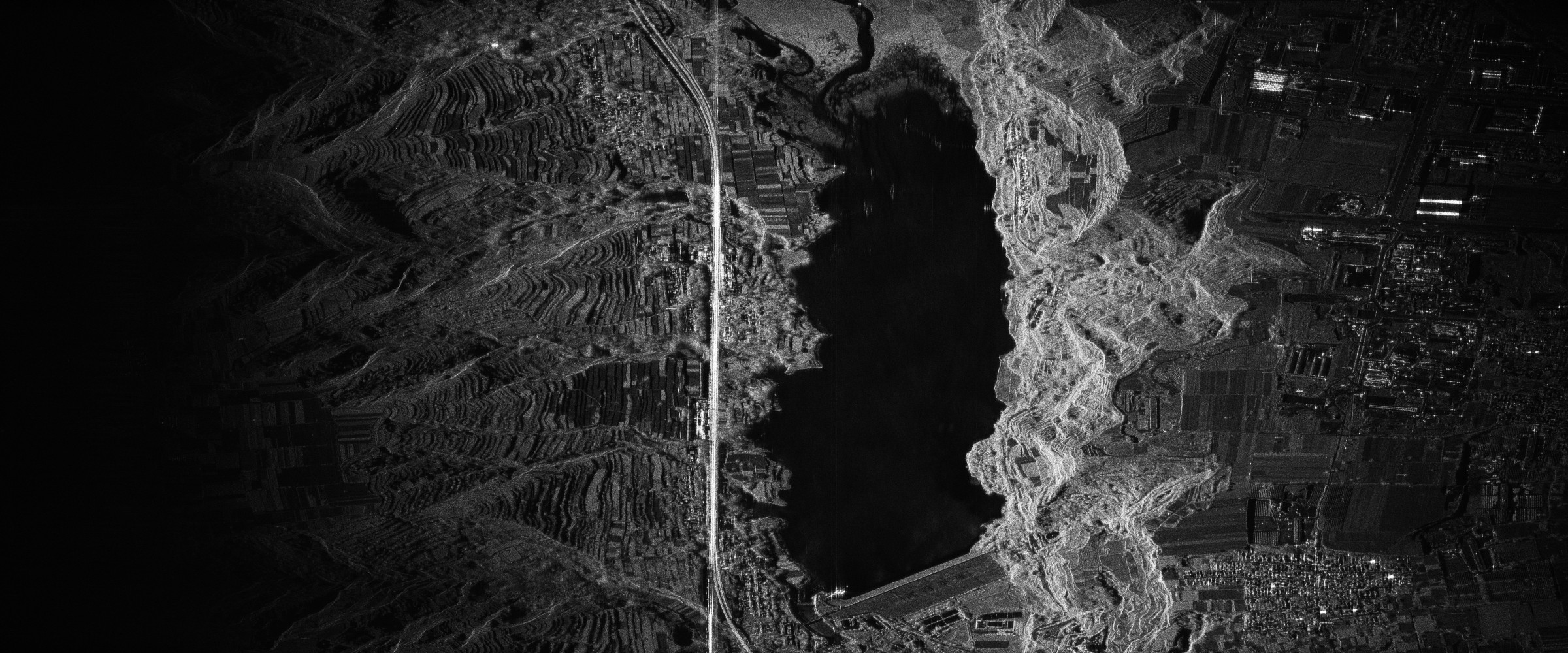}}
    \subfigure[]{\includegraphics[width=2.85cm,height=2.85cm]{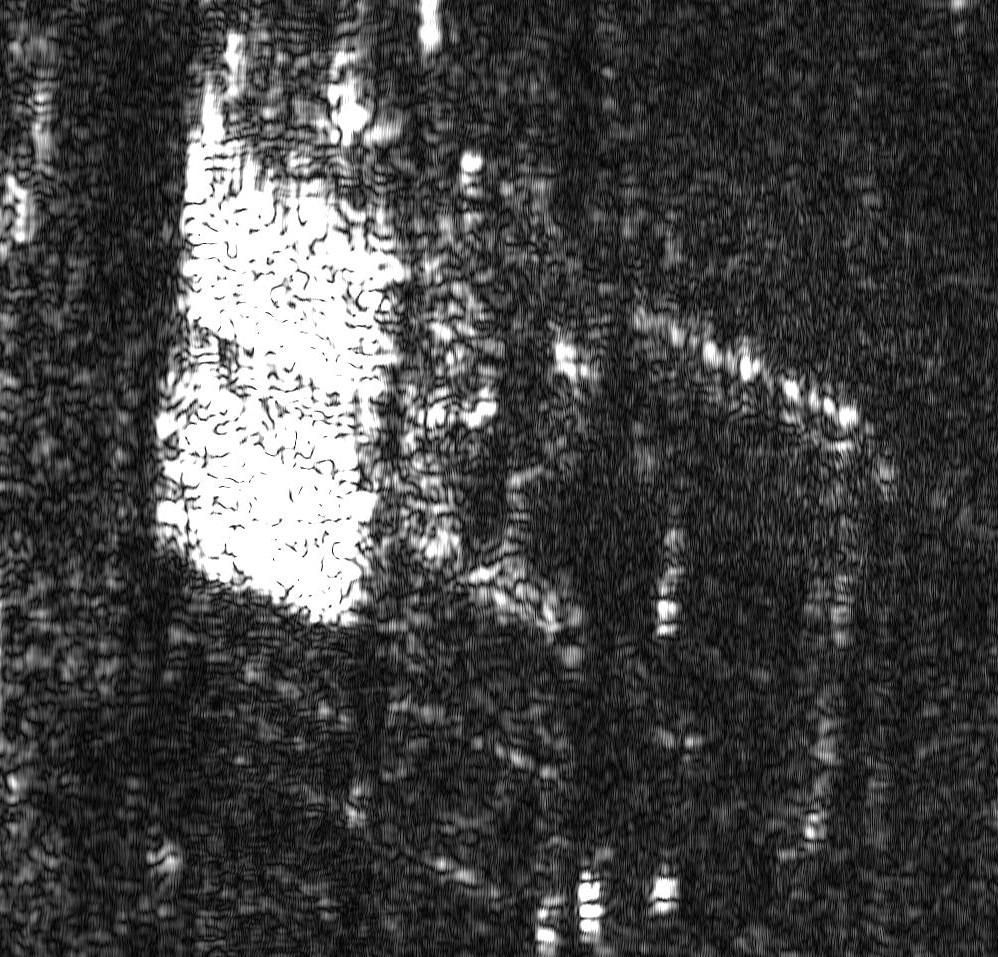}}
  \subfigure[]{\includegraphics[width=2.85cm,height=2.85cm]{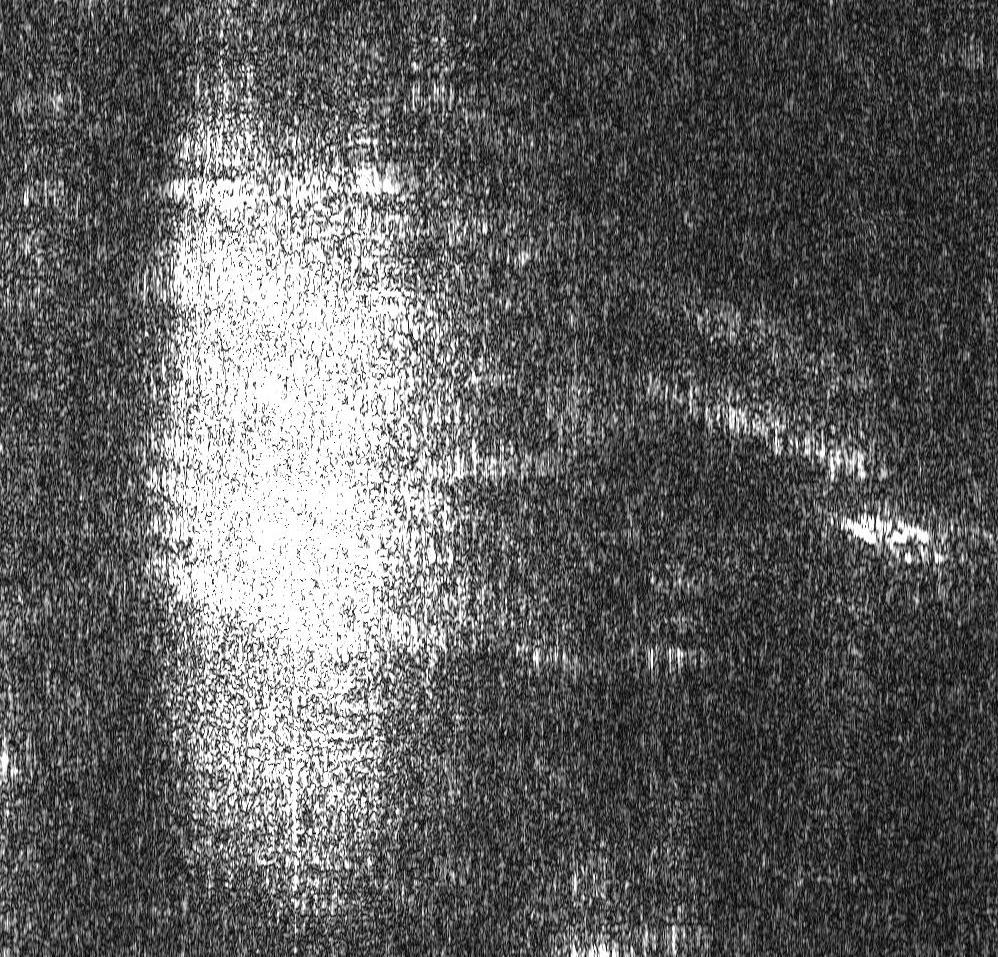}}
  \subfigure[]{\includegraphics[width=2.85cm,height=2.85cm]{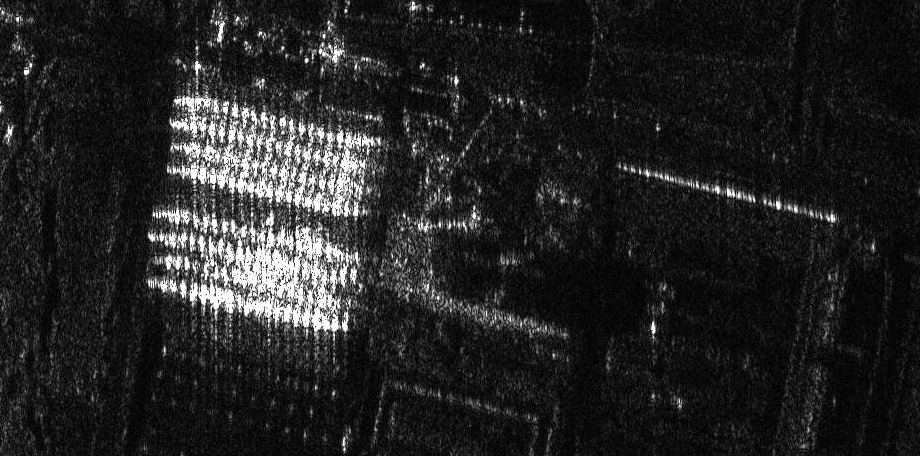}}
  \caption{The generated images in experiment \#6. (a) by method 1, (b) by method 2, (c) by method 3. (c)-(d) are detailed patches selected from (a)-(c), respectively.}\label{fig:e6}
\end{figure}

\subsection{Experiments \#2-\#4 using ERS Raw Data}

Following the first experiment, we use other three ERS raw data to perform further experimental comparison.

Specifically, in experiment \#2 we use  ERS data \url{E1_16687_STD_L0_F830}. The resulting images obtained by three methods are presented in Figs.  \ref{fig:e2}(a)-(c), respectively, and detailed single-look patches of the three images are shown in Figs.  \ref{fig:e2}(d)-(f), respectively. The imaged scene consists of two different areas: the left side is a land area and the right side is a sea area. The land area has higher backscattering coefficient than the sea, so the total clutter is significantly nonstationary. The method 2 fails to estimate an reference point echo under such clutter, leading to a low-quality image. On the contrary, the method 1 uses PCM across normalized segmented block to successfully estimate reference point echo and recover a high-quality image.

In experiment \#3 we use  ERS data \url{E2_14882_STD_L0_F337}. The resulting images obtained by three methods are presented in Figs.  \ref{fig:e3}(a)-(c), respectively, and detailed single-look patches of the three images are shown in Figs.  \ref{fig:e3}(d)-(f), respectively. The imaged scene consists of mountain area and some man made scatterers. The results show that method 1 generates a better image than method 2. From the detailed image patches, we can find that the image obtained by method 1 is also comparable to that obtained by the chirp scaling algorithm.

In experiment \#4 we use  ERS data \url{E2_11905_STD_L0_F357}. The resulting images obtained by three methods are presented in Figs.  \ref{fig:e4}(a)-(c), respectively, and detailed single-look patches of the three images are shown in Figs.  \ref{fig:e4}(d)-(f), respectively. The imaged scene consists of two different areas: the left side is mountain area and the right side is urban area.  The method 1 generates a well-focused image, but method 2 fails to generate a correct image. By checking Fig.  \ref{fig:e4}(b) and its detailed patches in Fig. \ref{fig:e4}(e), we can find that this image has severe range ambiguities. This behavior of method 2 is caused by the fact that the estimated reference echo of method 2 contains two points at the same azimuth cell but different range cells.

The above experiment results on ERS-1/2 data show that method 1 can effectively recover high-quality image from raw echo data, and is significantly robust compared to method 2.

\subsection{Experiment \#5 using RADARSAT-1 Raw Data}

In experiment \#5 we use  RADARSAT-1 data. The resulting images obtained by three methods are presented in Figs.  \ref{fig:e5}(a)-(c), respectively, and detailed single-look patches of the three images are shown in Figs.  \ref{fig:e4}(d)-(f), respectively. The raw data size is $1536\times2048$, and this data is part of the full raw data provided in the book ``Digital Processing of Synthetic Aperture Radar Data: Algorithms and Implementation" \cite{ cumming2005digital}.  The method 1 generates a well-focused image, but method 2 generates a blurred image. This failure of method 2 in this experiment is caused by the fact that the strong land clutter significantly deteriorates the estimation of a clean reference echo and thus causes an unfocused image.


\subsection{Experiment \#6 using Airborne Raw Data}

Different form the above experiments performed on spaceborne data, in experiment \#6 we use airborne data collected by the N-SAR system developed by the Nanjing Research Institute of Electronic Technology. The raw data size is $16000\times10000$ samples. We estimate the reference echo from its first $4000\times 10000$ samples to save time. The resulting images obtained by three methods are presented in Figs.  \ref{fig:e6}(a)-(c), respectively, and detailed single-look patches of the three images are shown in Figs.  \ref{fig:e6}(d)-(f), respectively. In this experiment, the third image obtained via extended chirp scaling algorithm with motion error compensation using GPS and inertial measurement unit data.  Comparing these obtained images, we find that method 1 can generate a focused image better than that obtained by method 2, but the image quality is not comparable to that obtained by the extended chirp scaling algorithm. A reason for this fact is the presence of motion error in the airborne data.

\begin{figure}
  \centering
  \subfigure[]{\includegraphics[width=4.3cm]{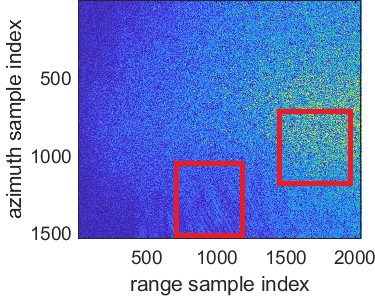}}\\
    \subfigure[]{\includegraphics[width=4.cm]{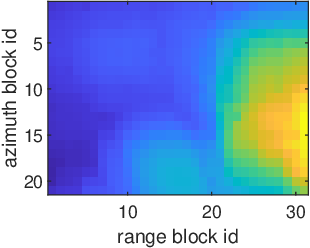}}
   \subfigure[]{\includegraphics[width=4.5cm]{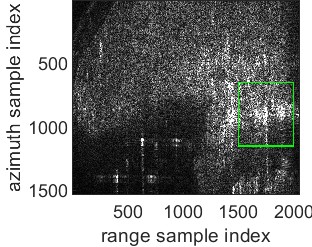}}\\
    \subfigure[]{\includegraphics[width=4.cm]{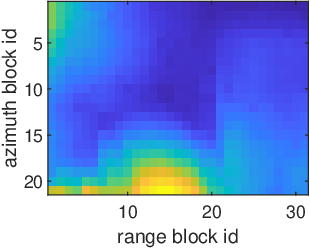}}
    \subfigure[]{\includegraphics[width=4.5cm]{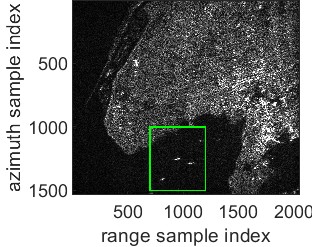}}
  \caption{Experimental results with and without the use of block normalization (\ref{eq:nm}). (a) the raw echo amplitude, (b) and (c) the maximum eigenvalues of all blocks and the generated image, without block normalization, (d) and (e) the maximum eigenvalues of all blocks  and the generated image, with block normalization.}\label{fig:e7}
\end{figure}

\begin{figure}
  \centering
  \subfigure[]{\includegraphics[width=4.3cm]{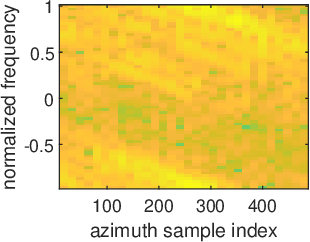}}
   \subfigure[]{\includegraphics[width=4.3cm]{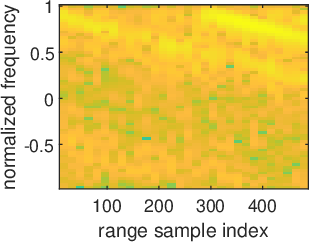}}\\
  \subfigure[]{\includegraphics[width=4.3cm]{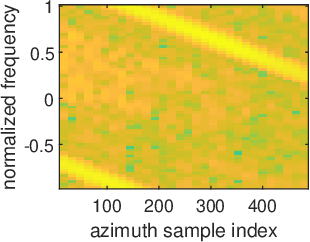}}
  \subfigure[]{\includegraphics[width=4.3cm]{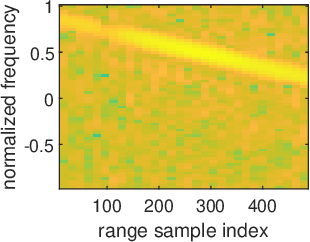}}
  \caption{The azimuth and range time-frequency spectrograms of the estimated reference echo from the RADARSAT data. (a) and (b) azimuth and range spectrograms obtained without block normalization, respectively. (c) and (d) those obtained with block normalization. }\label{fig:e7spg}
\end{figure}

\subsection{Analysis on the Effects of Block Normalization}

The key reason for the robustness of our method against clutter in the above experiments is the use of the block normalization, i.e., equation (\ref{eq:nm}). To better explain this point, we perform further experimental analysis using the same RADARSAT data described previously. By checking the image focused by chirp scaling algorithm in Fig. \ref{fig:e5}(c), we can find that the imaged scene consists of land urban area at the right side of the image and sea area at the left and bottom side of the image. In the bottom area of the image, there are several ships with strong point scatterers. Meanwhile, there are stronger point scatterers on the land urban area.
It is evident that the sea clutter is significantly weaker than that of the land urban area, which can be seen in the echo data in Fig. \ref{fig:e7}(a). Therefore, the strong point scatterer on the ship is a better choice as a reference point than the land point scatterer. However, if without block normalization (\ref{eq:nm}), the land point scatterer will be selected as the reference echo due to its associated echo data have stronger signal strength, which will lead to reduced image quality.

To verify the above mechanism, we generate images from this data using block size $500\times 500$ with and without block normalization. The adjacent blocks are set to have 450 overlapped samples, leading to total $21\times31$ blocks. The maximum eigenvalues of these blocks obtained with and without block normalization are shown in Figs. \ref{fig:e7}(b) and  \ref{fig:e7}(d), respectively. It is shown that, without block normalization, the reference echo is selected from the right side of the data that corresponds to land urban area in the image domain. On the contrary, with the use of block normalization, the reference echo is selected from the middle bottom area which corresponds to a ship's point scatterer. The resulting two images are shown in Figs. \ref{fig:e7}(c) and \ref{fig:e7}(e), respectively, and the areas associated with estimated echo are marked using green boxes. The results clearly show that the proposed method with block normalization generates a better image due to the successful estimation of reference echo from the ship's echo that has higher SCNR than the land point scatterer's echo.

We also provide the azimuth and range time-frequency spectrograms of the estimated reference echo in Figs. \ref{fig:e7spg}(a) and \ref{fig:e7spg}(b), where the Fig. \ref{fig:e7spg}(a) shows the spectrograms obtained without block normalization and Fig. \ref{fig:e7spg}(b) shows those obtained with block normalization.  It is evident that the latter results exhibit superior estimation of LFM signals compared to the former, which is why the latter yields a significantly cleaner image.


\section{Conclusion}\label{sec:conc}

This paper presents a novel approach for recovering SAR image from raw data without knowledge of SAR system parameters. 
We begin by establishing signal models for SAR echoes, encompassing point echo and its low-dimensional model as well as clutter model. These models provide a theoretical understanding of the underlying principles and characteristics of SAR signals.
Building upon the signal model, we introduce an approximate matched filtering model for SAR image formation. This model exploits the approximate shift-invariance of point echo responses, enabling image formation via matched filtering with an unknown reference echo.
To estimate the reference echo from raw 2-D data, we propose the PCM method. By segmenting the echo data into blocks and normalizing their energy, the PCM method effectively handles the impact of non-stationary clutter, leading to adaptive estimation of the reference echo.
A key aspect of the proposed method is the recovery of the 2-D raw data matrix from the 1-D raw data vector with an unknown pulse repetition interval (PRI). We present a two-step approach, utilizing the approximate periodicity of amplitude signals and PCM for coarse and fine PRI estimation, respectively. This enables the accurate transformation of 1-D data into a 2-D matrix for further processing.
To address the phase errors introduced by the approximate shift-invariant model, the paper proposes a range-varying azimuth reference signal estimation method. By compensating for the quadratic phase errors, the method ensures precise azimuth focusing and improves overall image quality.
The effectiveness and robustness of the proposed method are validated through extensive experiments using ERS, RADARSAT, and airborne SAR data. 

\bibliographystyle{IEEEtran}
\bibliography{refs0}

\end{document}